\documentclass[%
reprint,
superscriptaddress,
nofootinbib,
amsmath,amssymb,
aps,
pra, longbibliography
]{revtex4-2}

\usepackage{graphicx}
\usepackage{dcolumn}
\usepackage{bm}
\usepackage{bbm}
\usepackage[caption=false]{subfig}
\usepackage{braket}
\usepackage{amsmath}
\usepackage{lipsum}
\usepackage[T1]{fontenc}
\usepackage[utf8]{inputenc}

\begin{document}
\preprint{APS/123-QED}

\title{Performance of Rotation-Symmetric Bosonic Codes in a Quantum Repeater Network}

\author{Pei-Zhe Li}
\email{peili@nii.ac.jp}

\affiliation{School of Multidisciplinary Science, Department of Informatics, 
SOKENDAI (the Graduate University for Advanced Studies), 2-1-2 Hitotsubashi, Chiyoda-ku, Tokyo 101-8430, Japan}
\affiliation{National Institute of Informatics, 2-1-2 Hitotsubashi, Chiyoda-ku, Tokyo 101-8430, Japan}
\affiliation{Okinawa Institute of Science and Technology Graduate University, 1919-1 Tancha, Onna-son, Okinawa 904-0495, Japan}

\author{Josephine Dias}

\author{William J. Munro}

\affiliation{Okinawa Institute of Science and Technology Graduate University, 1919-1 Tancha, Onna-son, Okinawa 904-0495, Japan}

\author{Peter van Loock}

\affiliation{Institut f\"ur Physik
Johannes Gutenberg-Universit\"at Mainz,
Staudingerweg 7, 55128 Mainz, Germany}

\author{Kae Nemoto}

\affiliation{Okinawa Institute of Science and Technology Graduate University, 1919-1 Tancha, Onna-son, Okinawa 904-0495, Japan}
\affiliation{National Institute of Informatics, 2-1-2 Hitotsubashi, Chiyoda-ku, Tokyo 101-8430, Japan}

\author{Nicol\'o Lo Piparo}
\email{nicolo.lopiparo@oist.jp}

\affiliation{Okinawa Institute of Science and Technology Graduate University, 1919-1 Tancha, Onna-son, Okinawa 904-0495, Japan}


\begin{abstract}
Quantum error correction codes based on continuous variables play an important role for the implementation of 
quantum communication systems.
A natural application of such codes occurs within quantum repeater systems which are used to combat severe channel 
losses and local gate errors. In 
particular, channel loss drastically reduces the distance of communication between remote users. Here we consider a
cavity-QED based repeater scheme to address the losses in the quantum channel. This repeater scheme 
relies on the transmission of a specific class of rotationally invariant error-correcting codes. We compare several 
rotation-symmetric bosonic codes (RSBCs) being used to 
encode the initial states of two remote users connected by a quantum repeater network against the 
convention of the cat codes and we 
quantify the performance of the system using the secret key rate. In particular, we
determine the number of stations required to exchange a secret 
key over a fixed distance and establish the resource overhead.
\end{abstract}

\maketitle
\section{\label{intro}Introduction}
Long distance quantum communication can dramatically change the way users will exchange information in the near future
\cite{Duan2001,Jiang2007,Muralidharan2014}. The implementation of quantum networks 
\cite{Cirac1997,Chiribella2009,Perseguers2010} and a future quantum internet \cite{Kimble2008,Wehner2018,Munro2022} 
will rely on efficiently distributing entangled states over large distances. Such entangled states can then be 
used for various tasks such as distributed quantum computing \cite{OBrien2007}, quantum remote sensing 
\cite{Bandyopadhyay2000,Ikram2000} and quantum key distribution (QKD) \cite{Shi2001}. However, 
the required photons traveling through optical fibers obviously suffer from channel loss, making long-distance quantum communication challenging.
Quantum repeater (QR) systems \cite{Briegel1998,Duer1999,Loock2006,Jiang2009,Sangouard2011,Munro2015,Azuma2015,Nemoto2016}
are a natural solution that allows the distribution of entangled states ideally over distances ranging from a few kilometers to 
intercontinental distances. However, the implementation of such systems is still far from practical given the
state-of-the-art devices. In particular, quantum memories, a key building block of first generation 
of quantum repeaters \cite{Briegel1998,Duan2001}, are extremely demanding due to the long coherence times required. 

In the first generation QRs, entangled states are created over elementary links in a probabilistic but
heralded fashion and then entanglement purification is used to increase the quality of the Bell states \cite{Bennett1996}. After 
purification, entanglement swapping is then performed to extend the distance of the entangled pairs \cite{ifmmodeZelseZfiukowski1993,Pan2001}.
Due to the heralded
nature of both procedures, classical communication is required to relay the success of those operations \cite{Briegel1998,Sangouard2011}. 
In second generation QRs, only the entanglement distribution step is done in such a heralded fashion. Error correction 
protocols are applied to correct operational errors (excluding loss) \cite{Jiang2009,Munro2010,Zwerger2014}. Finally, in the third generation QRs, 
quantum error correction protocols are applied to address both loss and operational errors. These repeater 
protocols do not require two-way classical signaling meaning in principle, they can achieve much faster distribution rates \cite{Fowler2010,Munro2012,Muralidharan2014,Azuma2015,Muralidharan2016}. 
However, the realization of efficient quantum error correction protocols is extremely challenging and requires significant physical resources, making third generation QR protocols hard to achieve in the near future.

A recent third generation QR scheme has been proposed in \cite{Li2023}, based on encoding quantum information into photonic logic states which are part of the rotation-symmetric bosonic codes (RSBCs) \cite{Grimsmo2020}. 
Such codes are invariant upon a rotation of a certain angle in phase space. 
Further like all third generation schemes, it does
not require quantum memories since all the operations are made near instantaneously \cite{Fowler2010,Munro2012}. 
In \cite{Li2023}, the QR's performance using RSBC group cat states was established, 
but it
required the elementary distance between stations to be around 0.1 km.
In the current work, we determine the performance of such repeater schemes 
using other types of RBSCs, such as the squeezed cat \cite{Liu2005,Schlegel2022,xu2022b}, binomial \cite{Michael2016}, 
Pegg-Barnett \cite{Grimsmo2020} and GKP-like codes
\cite{Ft1,Ft2}. We compare these codes calculating 
the SKR per channel and show that, generally, the elementary distance can be 
extended. Then for the best performing code, we evaluate
the necessary resource overhead and show that this repeater scheme is similar in performance to existing third
generation repeater schemes.

The paper is structured as follows: In Sec. \ref{rsbc} we introduce the rotation-symmetric bosonic codes describing the main features
analyzed in this work. Next in Sec. \ref{pro} we review the memoryless repeater scheme and the steps 
required to generate a secret key between remote users.
Then in Sec. \ref{skr} we present the main figure of merit used to assess the performance of the quantum repeater protocol 
followed in Sec. \ref{Ana} by our comparison of the RSBCs. Finally in Sec. \ref{Con} we summarize our results.

\section{\label{rsbc}Rotation-Symmetric Bosonic Codes}
Let us begin by reviewing the main properties of the quantum error correction codes we will utilize in our work. These belong to the set of 
the rotation-symmetric bosonic codes \cite{Grimsmo2020}. 
A quantum error correction code is
M-fold rotationally symmetric if it is 
invariant under the operation of the discrete rotation operator $\hat{R}(2\pi/M)=e^{i(2\pi/M)\hat{n}}$, 
which rotates the state by the angle $2\pi/M$ in phase space. 
The logic states of the RSBCs are given by \cite{Li2023}
\begin{eqnarray}
    \label{eq1}
    \begin{aligned}
        &\left|\left.0_{M,\Theta}\right\rangle\right.=\frac{1}{\sqrt{N_M}}\sum_{k=0}^{M-1}{\hat{R}\left(\frac{2k\pi}{M}\right)\left|\Theta\right\rangle},\\
        &\left|\left.1_{M,\Theta}\right\rangle\right.=\frac{1}{\sqrt{N_M}}\sum_{k=0}^{M-1}{\hat{R}\left(\frac{\left(2k+1\right)\pi}{M}\right)}\left|\Theta\right\rangle,
    \end{aligned}
\end{eqnarray}
where $N_M$ is a normalization constant while $\left|\Theta\right\rangle$ is the primitive state, which 
can be any arbitrary optical state with the constraints that it is composed of at least one of $\ket{2kM}$ Fock states and one of 
$\ket{(2k+1)M}$ Fock states, respectively, where $k\in\mathbb{N}$ \cite{Grimsmo2020}. This code can tolerate errors caused by $l\leq M-1$ discrete
photon losses. Here $l$ is defined as the loss order of the code \cite{Li2023}. 
It is important to mention that the codewords defined above are not necessarily orthogonal meaning the discrimination of them may not be deterministic.
That in turn means that the error correction procedure can only be realized in a heralded but probabilistic fashion
in a protocol requiring state discrimination. 
In the following subsections, we will describe the main features of the error correction codes we analyze in this work beginning with the squeezed cat one.

\subsection{\label{squd}Squeezed cat codes}
The squeezed cat codes are an extension of the cat codes \cite{Mirrahimi2014,Bergmann2016,Leghtas2013} with 
its primitive state $\left|\Theta\right\rangle$ given by
\begin{eqnarray}
    \label{eq2}
    |\alpha,-r\rangle=\hat{D}(\alpha)\hat{S}(-r)|0\rangle,
\end{eqnarray}
where $\hat{D}(\alpha)=\exp(\alpha\hat{a}^{\dagger}-\alpha^*\hat{a})$ is the 
displacement operator while $\hat{S}(r)=\exp[1/2(r\hat{a}^{\dagger2}-r^*\hat{a}^{2})]$ is the squeezing operator \cite{Stoler1970,Lu1972,Hollenhorst1979}. 
For convenience, we will assume both $\alpha$ and $r$ to be real and positive. 
Our logic states of the squeezed cat code are given by \cite{Ft3}
\begin{eqnarray}
    \label{eq3}
    \begin{aligned}
        &\left|\left.0_{M,sc}\right\rangle\right.=\frac{1}{\sqrt{N_{M,sc}}}\sum_{k=0}^{M-1}{\hat{R}\left(\frac{2k\pi}{M}\right)|\alpha,-r\rangle},\\
        &\left|\left.1_{M,sc}\right\rangle\right.=\frac{1}{\sqrt{N_{M,sc}}}\sum_{k=0}^{M-1}{\hat{R}\left(\frac{\left(2k+1\right)\pi}{M}\right)|\alpha,-r\rangle},
    \end{aligned}
\end{eqnarray}
where $N_{M,sc}$ is the appropriate normalizaton constant.
\subsection{\label{bid}Binomial codes}
Next another interesting RSBC code is the binomial code defined as a finite superposition of Fock states \cite{Michael2016} 
with coefficients that follow a binomial distribution. Its logic states are defined as
\begin{eqnarray}
    \label{eq4}
    \begin{aligned}
        &\left|\left.0_{M,bin}\right\rangle\right.=\frac{1}{2^{K/2}}\sum_{k=0}^{K}{\sqrt{\binom{K}{k}}|kM\rangle},\\
        &\left|\left.1_{M,bin}\right\rangle\right.=\frac{1}{2^{K/2}}\sum_{k=0}^{K}{(-1)^k\sqrt{\binom{K}{k}}|kM\rangle},
    \end{aligned}
\end{eqnarray}
where $K=1,2,...$ sets the truncation level.

\subsection{\label{gkpd}GKP codes}
The Gottesman, Kitaev, and Preskill (GKP) codes are considered to have the highest error correction performance based on the discrete 
translation symmetry in phase space \cite{Gottesman2001}. 
The original GKP states have 2-fold rotation symmetry and some approximate GKP codes 
such as the square- and hexagonal-lattice GKP codes have even higher rotation symmetries but cannot exceed 6-fold \cite{Grimsmo2020}.
They, however, do not belong to the set of RSBC. 
Here we construct a so-called GKP-like code which belongs to the set of RSBC defined as follows \cite{Albert2018}:
\begin{widetext}  
    \begin{eqnarray}
        \label{eq6}
        \begin{aligned}
            \ket{0_{M,gkp}}=&\frac{1}{\sqrt{N_{M,gkp}}}\sum_{k=0}^{M-1}{[\hat{R}(\frac{2k\pi}{M})(e^{-\Delta^2\alpha^2}\ket{\alpha}+e^{-4\Delta^2\alpha^2}\ket{2\alpha})]},\\
            \ket{1_{M,gkp}}=&\frac{1}{\sqrt{N_{M,gkp}}}\sum_{k=0}^{M-1}{[\hat{R}(\frac{(2k+1)\pi}{M})(e^{-\Delta^2\alpha^2}\ket{\alpha}+e^{-4\Delta^2\alpha^2}\ket{2\alpha})]},
        \end{aligned}
    \end{eqnarray}
\end{widetext} 
where $\Delta>0$ is a parameter that determines the weight balance between the terms \cite{Albert2018}.
As usual, $N_{M,gkp}$ is the normalization constant. 
\begin{figure}
    \includegraphics[width=7.9cm]{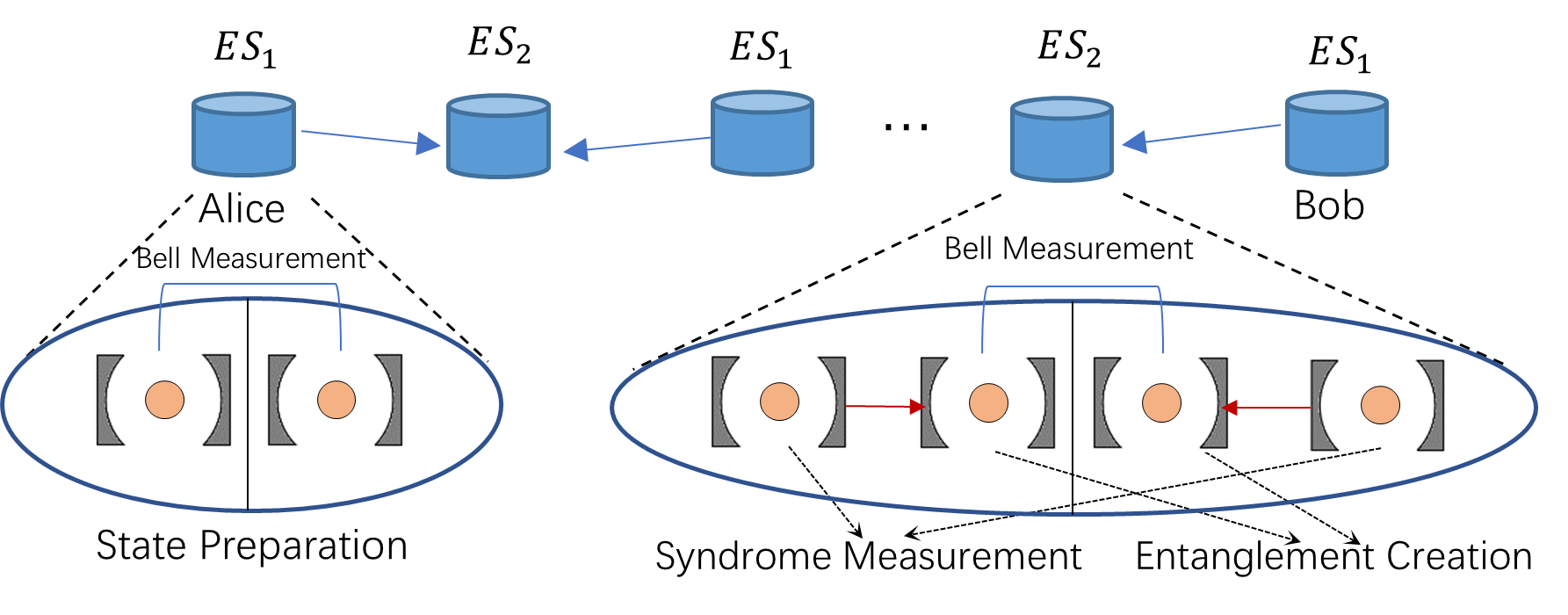}
    \captionsetup{justification=raggedright, singlelinecheck=false}
    \caption{\label{figrs} Schematic illustration of our long-distance 
    entanglement distribution protocol including two types 
    of elementary stations: $ES_1$ transmits the light mode across the channel
    to $ES_2$ which receive it. $ES_1$
    consists of two cavities which are used to 
    generate the desired light mode state and the two modes 
    just generated are then sent to the two nearest 
    stations  separately. For $ES_2$, the cavities are used to perform the syndrome 
    measurement and to establish the entanglement. The corresponding light-matter interaction is described by
    $U_\phi=\exp(i\phi \ket{\downarrow}\bra{\downarrow}\otimes \hat{n})$.
    }
\end{figure}

\section{\label{pro}Protocol}
In this section we provide a short description of the quantum repeater protocols we will use later. A more detailed description can be found in \cite{Li2023}.
It is well-known that in a general quantum repeater protocol the quantum channel connecting two remote users, Alice and Bob, separated by a distance $L_{tot}$, is divided into smaller segments over which entangled pairs are created.
Then, through a series of entanglement swapping operations, the entangled pairs can be extended until they reach Alice and Bob. 
The quantum repeater protocol analyzed in this work follows a similar architecture of the general case and it can be divided into three main steps. 
First, 
state preparation is performed in which a photonic state 
is entangled with the energy levels of a trapped atom and sent through an elementary link from the station $ES_1$ (see Fig. \ref{figrs}). 
Then, at the station labelled $ES_2$ in Fig. \ref{figrs}, two incoming photonic states go through the syndrome measurement step to increase the fidelity of the pairs distributed over the elementary links.
After the syndrome measurement, Bell-state measurements (BSMs) are performed simultaneously between the cavities of each station $ES_1$ and $ES_2$, respectively. Upon successful BSMs, Alice and Bob will eventually share an entangled pair. 
Next we will section how we plan to quantify the performance of such quantum repeaters. 

\section{\label{skr}secret key rate}
In this section, we use the formula of the SKR under the assumption that Alice and Bob perform the
BBM92 protocol \cite{Bennett1992}.
The SKR would then be given by $R_{QKD}:=R_{raw}r_{\infty},$
where $R_{raw}$ is the raw key rate while $r_{\infty}$ is the 
secret fraction \cite{Scarani2009}. In a memoryless 
repeater scheme, the raw key rate can be written as $R_{raw}=P_{tot}/t_0,$
where $P_{tot}$ is the total success probability and $t_0$ is the repetition time. In the quantum repeater protocol associated with Fig. \ref{figrs}, 
the repetition time $t_0$ depends on the slowest operation, namely the initialization/detection time of the trapped atoms used in the elementary stations, assuming that the interaction time between the photons and the cavities is much smaller.
Next the secret fraction is given by 
$r_{\infty}=1-h(e_z)-h(e_x)$,
where $h(p)$ is the binary entropy $h(p)=-p\log_{2} p-(1-p)\log_{2} (1-p)$ with ($e_x$, $e_z$)
being the quantum bit error rates (QBERs) \cite{Abruzzo2013,Kirby2016}. In this situation, we can substitute $e_x$ and $e_z$ with the total fidelity $F_{tot}$ \cite{Li2023}.
Then the total key 
rate per channel use, $\mathcal{R}$, is lower bounded by
\begin{eqnarray}
    \label{eq10}
    \mathcal{R}=t_0R_{QKD}>\mathcal{R}_{LB}=P_{tot}(1-h(F_{tot})).
\end{eqnarray}
We remark that the codewords of the RSBCs are not perfectly 
orthogonal, so the discrimination of the codewords can not 
be deterministic and will result in a success probability 
which is smaller than one within a single unit. 
Since in our quantum repeater system we are not using any quantum memories, the total probability is given by the product 
of the success probabilities of each elementary link $P_0$, that is, $P_{tot}=(P_0)^n$, where $n$ is the number of elementary links.
The success probability of each elementary link $P_0$ is the probability to distinguish 
the two codewords after channel loss. This is given by $P_0=1-|\braket{\tilde{0}_{M,\Theta}|\tilde{1}_{M,\Theta}}|$ when unambiguous state
discrimination is used \cite{Li2023}. Above $\ket{\tilde{0}_{M,\Theta}}$ and $\ket{\tilde{1}_{M,\Theta}}$
are the damped codewords after the transmission of the states over the elementary link.
\section{\label{Ana}Performance}

Let us now compare the performance of the repeater protocol of Fig. \ref{figrs}, when the information is encoded into 
squeezed cat codes, binomial codes and GKP-like codes,
respectively \cite{Ft3}. When such codes are used, we will establish the SKR defined in Eq. \ref{eq10}. 
The performance of this repeater scheme based on cat codes was analyzed in
\cite{Li2023}, so we will benchmark our RSBCs against them.

\subsection{\label{squ}Squeezed cat codes}

\begin{figure}
    \includegraphics[width=7.9cm]{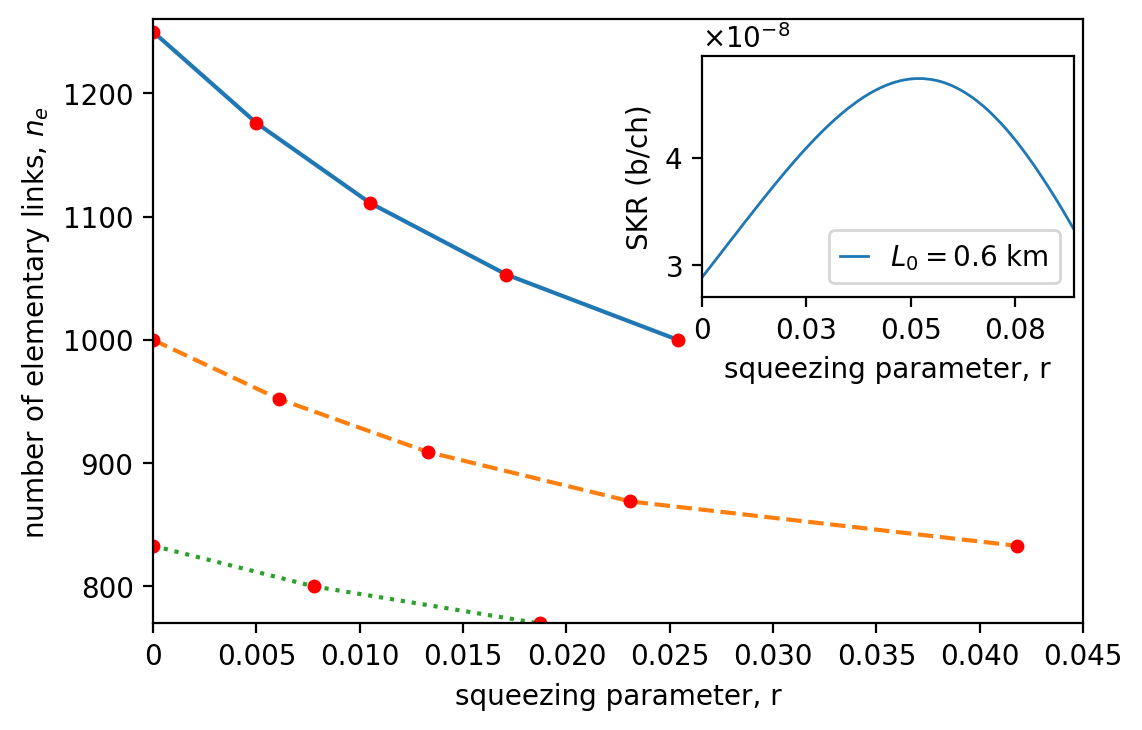}
    \captionsetup{justification=raggedright, singlelinecheck=false}
    \caption{\label{figsq}
    The resources ($n_e$) required to  
    reach the target SKR with respect to $r$ for total distance $L_{tot}=500$ km.
    All data points on the same curves achieve the same target SKR. 
    The target SKR of the points on the green (solid), orange (dashed) and blue (dotted) curves equals 
    the optimized SKR of the cat codes ($r = 0$) with elementary distance $L_0=0.4$ km, $L_0 = 0.5$ km 
    and $L_0 = 0.6$ km, respectively.    
    By increasing the squeezing, 
    less resources are required to achieve the same key rate.
    The inset figure shows the SKR in bits per channel use (b/ch) for the squeezed cat codes with respect to $r$ for
    $L_{tot}=500$ km with an elementary distance $L_0=0.6$ km.
    }
\end{figure}

Let us begin with the squeezed cat codes where we initially determine the value of $\alpha$ that optimizes the SKR for the 1-loss squeezed cat and 1-loss cat codes. This depends on $L_0$, $L_{tot}$ and $r$.  
Next we truncate the squeezed state at $N=20$ Fock state since 
the SKR is always optimized when $\alpha<3$.
The inset of Fig. \ref{figsq} shows the SKR versus the squeezing parameter $r$ when $L_0=0.6$ km with $L_{tot}=500$ km. For $0<r<0.1$, we observe that the SKR of the squeezed cat code is higher than the one of the cat code ($r=0$). This shows that using squeezed cat states can bring an advantage compared to standard cat states. 
For instance, when the states are encoded into the cat code, at a total distance $L_{tot}$ = 500 km, we can reach a key rate  $\mathcal{R}=7.48\times 10^{-8}$ b/ch using 1250 elementary links (corresponding to $L_0=0.4$ km) optimized over the value of $\alpha$, as shown in the blue curve of Fig. \ref{figsq}.
On the other hand, when the squeezed cat codes are used, the elementary distance can be increased while keeping similar SKR to the cat states encoding. In particular, for a modest squeezing of $r = 0.025$, the number of stations is $20\%$ less when compared with cat codes for $L_0=0.4$ km. 
The same advantage is obtained for $L_0= 0.5$ km (red curve) and $L_0 = 0.6$ km (green curve), and is valid for $L_0$ less than $0.8$ km. Higher values of $r$ do not reduce the number of elementary links further.

\subsection{\label{bi}Binomial codes}
\begin{figure}[]
    \includegraphics[width=7.9cm]{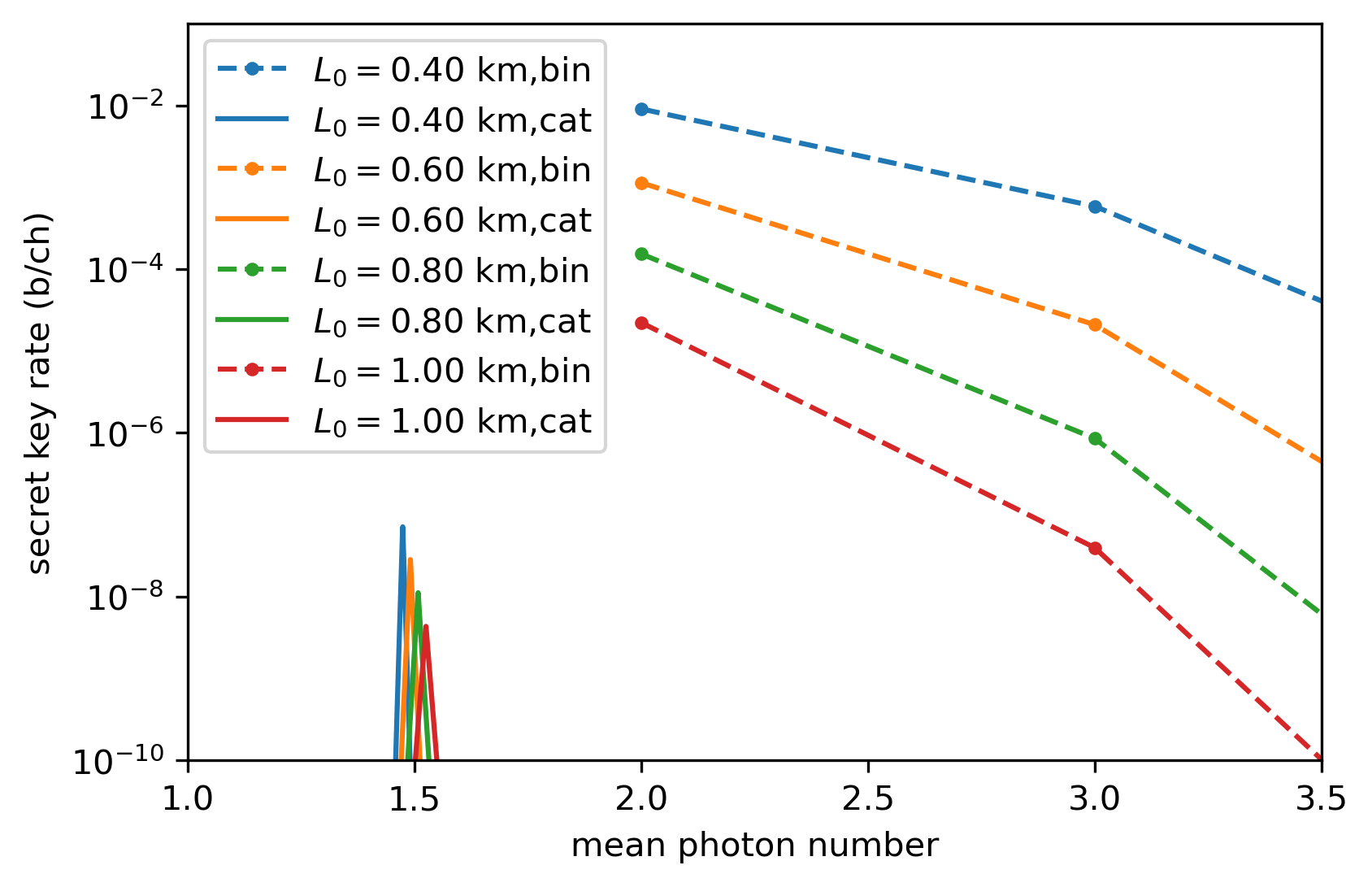}
    \captionsetup{justification=raggedright, singlelinecheck=false}
    \caption{\label{figbi}The SKR for 1-loss cat and binomial codes for various elementary distance $L_0$ ranging from $0.4$ km to $1$ km.}
\end{figure}

Next, let us consider our second encoding, the binomial codes. Figure \ref{figbi} shows the comparison between the SKR of the cat code and binomial code versus the mean photon number for loss order $l = 1$ at various elementary distances.
The SKR of the repeater scheme with binomial codes is much higher, 
as shown in Fig. \ref{figbi}. For instance, at $L_0 = 1$ km, the SKR of the binomial code 
(red dashed line in Fig. \ref{figbi}) is optimized to be $10^{-5}$ b/ch for a mean photon number of $2$ 
whereas the cat codes is optimized to be around $10^{-8}$ b/ch for a mean photon number to be near $1.5$. 
As $L_0$ decreases, the SKRs of both the cat and binomial codes become higher but the SKR for binomial codes is always much higher than cat codes.
The SKR curves in Fig. 3 for the binomial codes are derived numerically by 
assuming that the density matrix obtained when $q+k(l+1)$ photons are lost is proportional to the density matrix obtained for the loss of $q$ photons, where $q\in \mathbb{N}, q\leq l$, 
$\forall k\in \mathbb{N}, k\geq 1$, which is the actual case for the cat codes. 
Based on this assumption the SKRs of Fig. \ref{figbi} represent upper bounds of the actual rates. In a more realistic scenario we assume that the density matrix obtained after the channel loss to be proportional also to a complete mixed state with some probability, which has been estimated in Appendix \ref{apx}.
The final result shows that the upper bounds in Fig. \ref{figbi} are not too different from the accurate value.
Thus, the comparison 
here between the SKRs of binomial codes and cat codes is still valid.

\subsection{\label{gkp}GKP-like codes}
\begin{figure}
    \includegraphics[width=7.9cm]{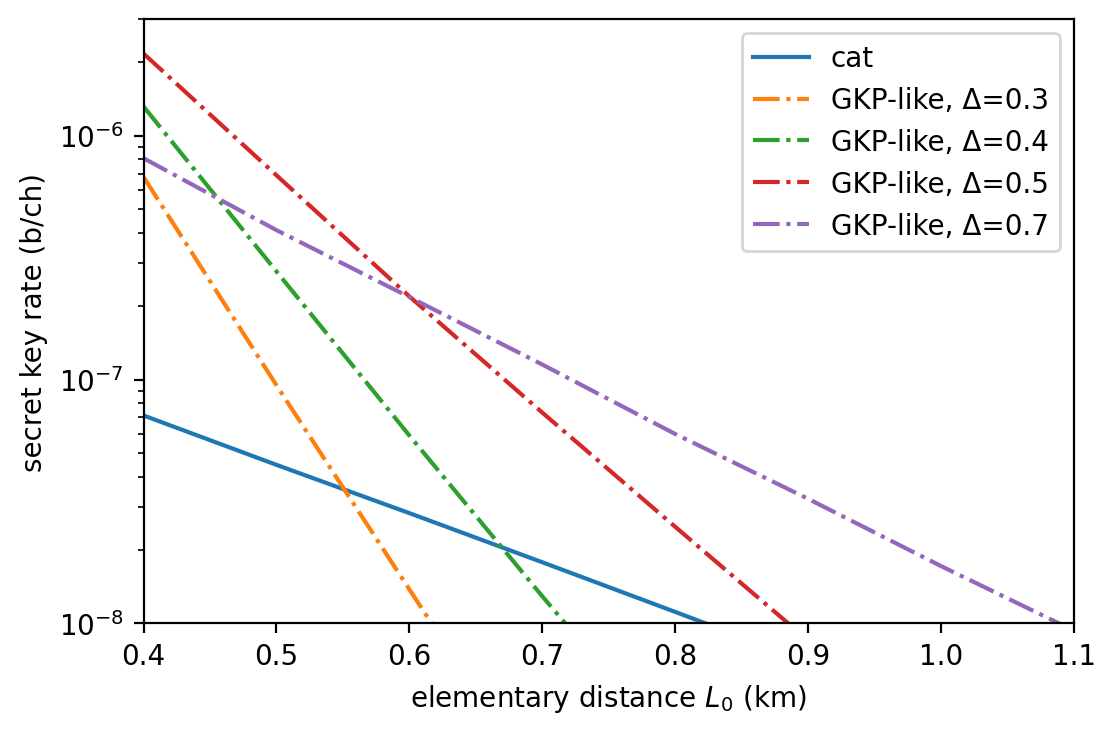}
    \captionsetup{justification=raggedright, singlelinecheck=false}
    \caption{\label{figgkp}The optimized SKR over $\alpha$ for the 1-loss GKP-like and cat codes versus elementary distance $L_0$ for a total distance $L_{tot}=500$ km.}
\end{figure}

Our final code is considered here the GKP-like codes. The comparison of SKR for 1-loss squared GKP-like code 
with cat code is shown in Fig. \ref{figgkp}, where the optimized SKR over the value of mean photon number for
1-loss GKP-like code with certain values of $\Delta$ can be an order of magnitude larger than that for 1-loss cat code with the same 
elementary distance ($L_0=0.4$ km), the SKR for GKP-like code with $\Delta=0.3$
is $\sim7\times10^{-6}$ b/ch while the SKR for cat code is $\sim7\times10^{-7}$ b/ch. Also, for a 
certain desired SKR, the corresponding elementary distance using the 1-loss GKP-like code can be larger
than that of the 1-loss cat code, thus, reducing the number of repeater stations. For instance, at $\mathcal{R}=10^{-8}$ b/ch,
the elementary distance for the 1-loss GKP-like code with $\Delta=0.7$ is near 
$1.1$ km while the elementary distance for 1-loss cat code is near $0.82$ km.
As shown in Fig. \ref{figgkp}, the changing of the parameter $\Delta$ affects the SKR significantly.
As $\Delta$ increases, the magnitude of the slope of the curve for the SKR of the GKP-like code becomes smaller and the curve will 
eventually overlap with that of the cat code. 

\begin{figure}
    \subfloat[]{
        \includegraphics[width=0.46\textwidth]{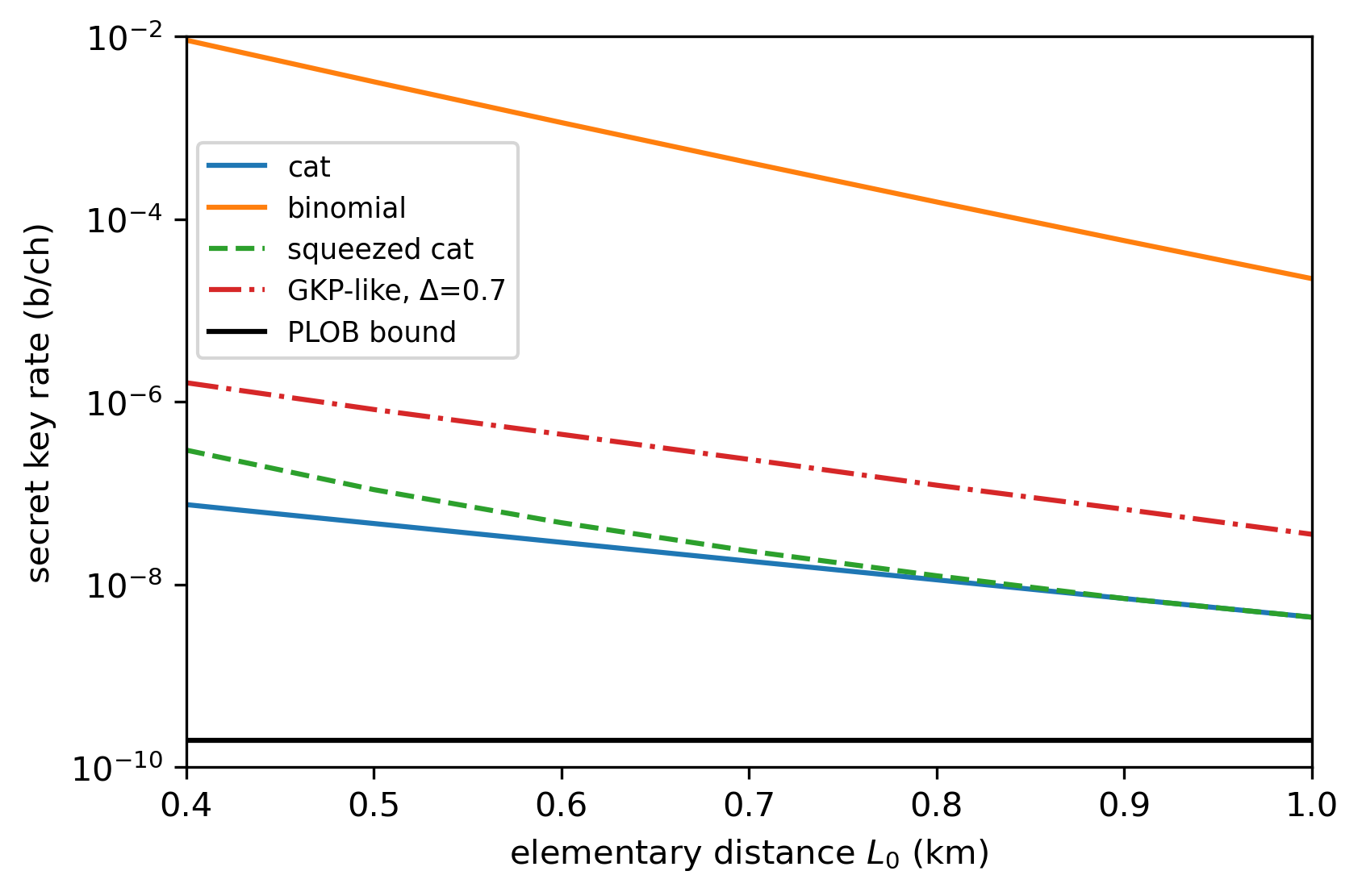}
        \label{rsbcc}}
    \hfill 
    \subfloat[]{
        \includegraphics[width=0.46\textwidth]{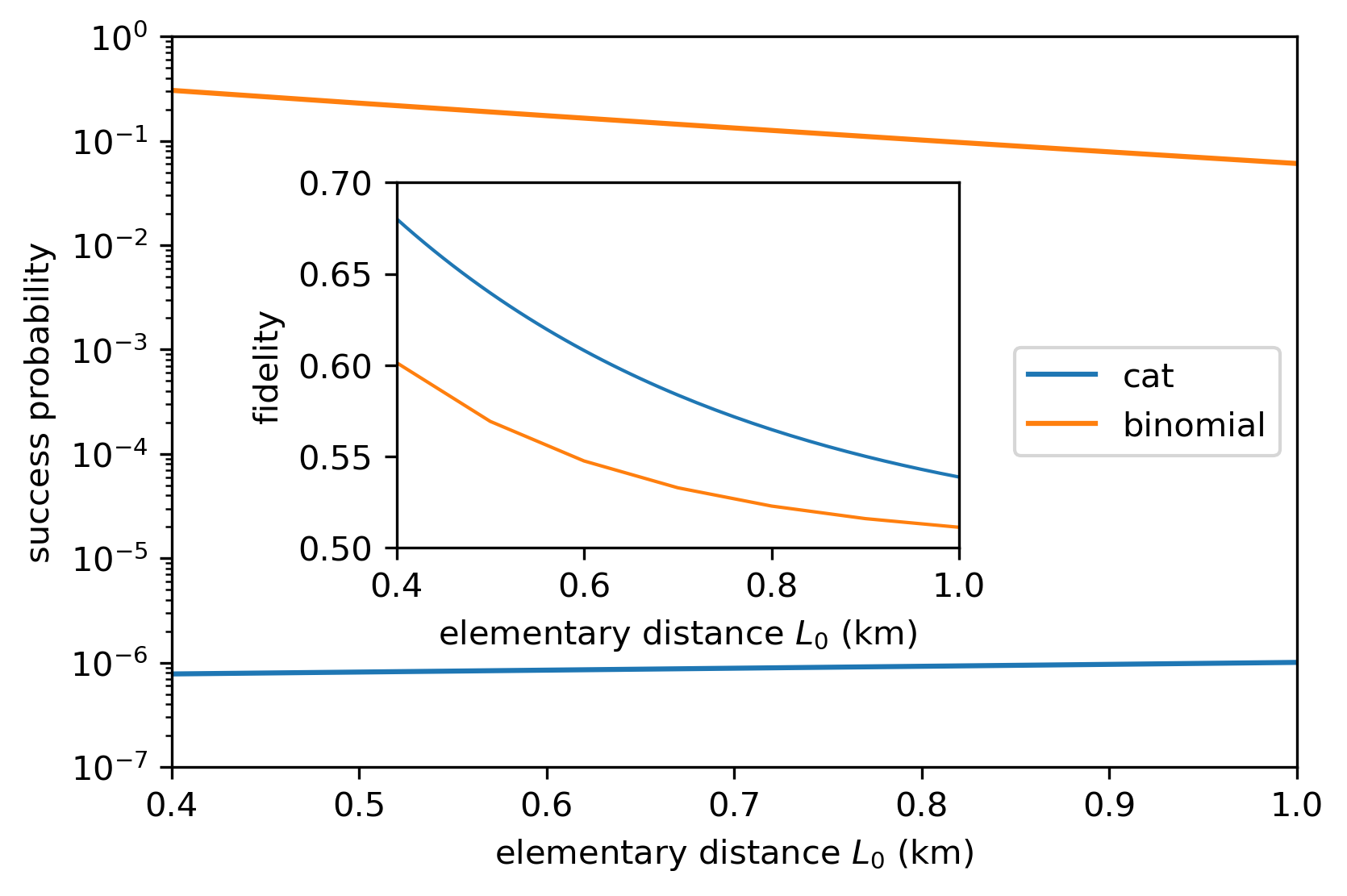}
        \label{binf}}
    \caption{The optimized SKR for 1-loss GKP-like codes, binomial, squeezed cat and cat codes versus elementary distance is shown in (a).
        The SKR is optimized over the value of mean photon number (and squeezing parameter $r$) for these codes.
        The fidelity and success probability of the 1-loss binomial codes and cat codes versus elementary distance with the same optimized values of mean photon number as in (a)
        are shown in (b).
        The total distance $L_{tot}$ is fixed to be $500$ km.}\label{figrsbc}
\end{figure}

\subsection{\label{cop}Performance comparison}

So far we have compared a number of different RSBCs to the cat codes. It is essential now to compare them among themselves. Figure \ref{figrsbc} shows the comparison of the SKRs of the codes we have discussed so far versus the elementary distance, $L_0$ for loss-order equal to $1$ km and $L_{tot} = 500$ km.
The SKR of the
squeezed cat codes (green dashed line) is higher than the one of the cat codes at $L_0$ less than around $0.8$ km.
The SKR of the GKP-like codes is higher than the one of the squeezed cat codes with appropriate 
values of $\Delta$, e.g., the SKR of the GKP-like codes is always higher than the one 
of the squeezed cat codes with $0.4$ km$<L_0<1$ km for $\Delta=0.7$. 
Figure \ref{figrsbc} shows that the SKR of the binomial codes outperforms largely the SKRs of the rest of the codes, reaching $4$ orders of magnitude higher than the one of the GKP-like code.
This behavior can be explained by looking at the comparison of the fidelity 
(see Fig. \ref{binf}) and the 
success probability (see the inset of Fig. \ref{binf}) of the binomial codes and the cat codes 
versus elementary distance.
Figure \ref{binf} shows that the fidelity of the binomial codes is slightly higher than
the fidelity of the cat codes at various elementary distance. 
However, in the inset of Fig. \ref{binf}, one can see that the success probability of the 
binomial codes is much higher than that of the cat codes. This is due to the fact 
that the codewords of the binomial code are initially orthogonal. The channel 
loss affects the orthogonality of such states only in a minimal way, therefore, the overlap 
of the codewords is only slightly modified after the transmission. This in turn, 
will increase tremendously the probability of success $P_0$.
\subsection{\label{pco}Performance cost}
\begin{figure}
    \includegraphics[width=7.9cm]{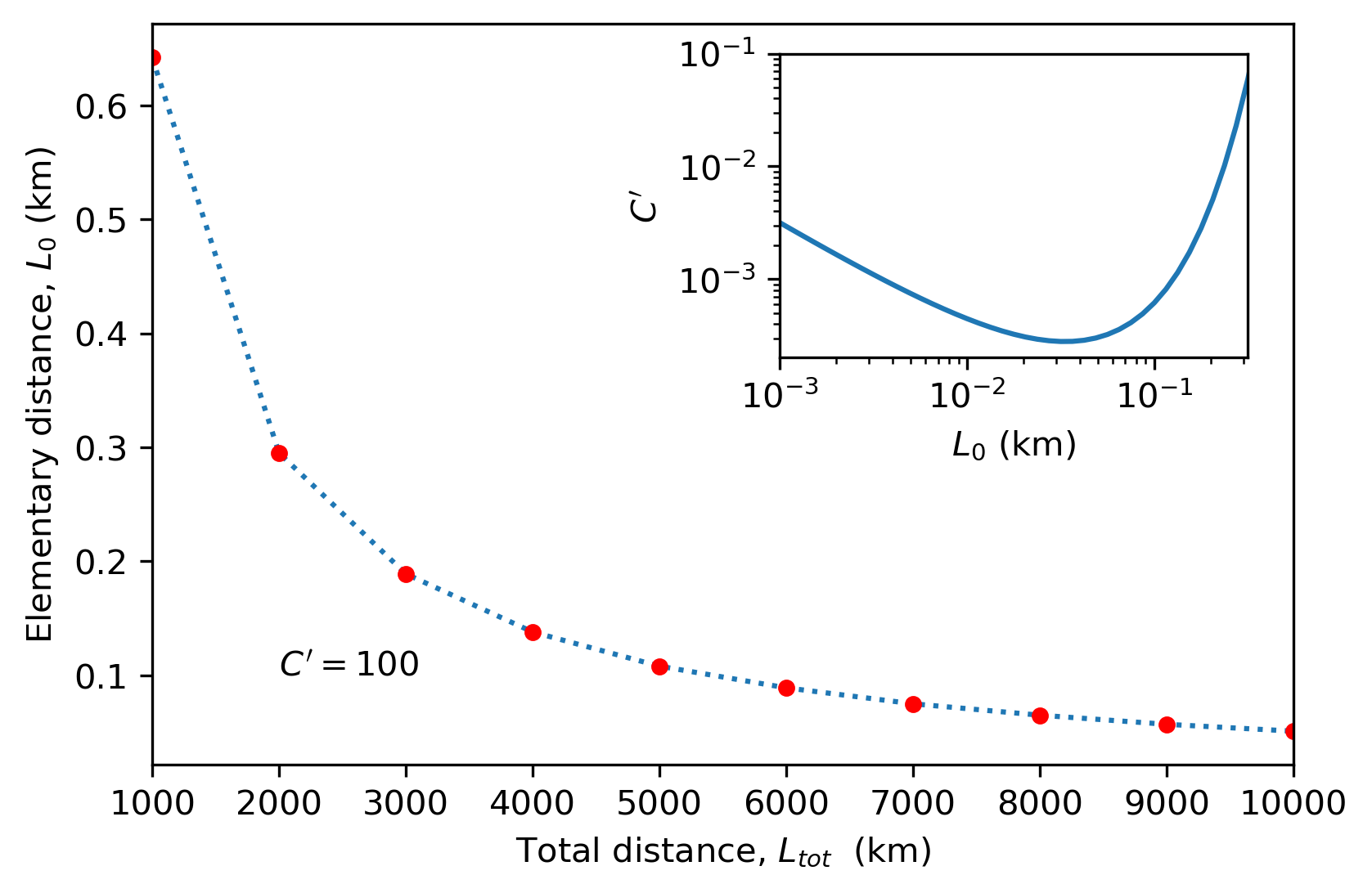}
    \captionsetup{justification=raggedright, singlelinecheck=false}
    \caption{\label{figcf}The elementary distance $L_0$ required to reach the cost coefficient value $C^\prime=100$ versus the total distance $L_{tot}$ for 1-loss binomial code.
    The inset presents $C^\prime$ versus $L_0$ (for $L_{tot}=1000$ km) and shows the lowest achievable $C^\prime$.}
\end{figure}

To evaluate the resource overhead for our scheme, we calculate the cost coefficient $C^\prime$ defined by $C^\prime=C/L_{tot}$, where the cost function $C$ is given by \cite{Muralidharan2014}
\begin{eqnarray}
    \label{eqcf}
    C=\frac{N_{tot}}{R_{QKD}}=\frac{N_s}{R_{QKD}}\times\frac{L_{tot}}{L_0}.
\end{eqnarray}
Here, $N_{tot}$ is the total number of matter qubits used in the repeater network while
$N_s$ is the number of matter qubits within each elementary link. In the situation here, $N_s$ is
the number of trapped atoms/ions used to establish the
entangled pairs between adjacent repeater stations (distance $L_0$) and perform the syndrome
measurement. Figure \ref{figcf} illustrates the elementary distance
required to achieve a fixed cost coefficient ($C^\prime=100$) using 
the 1-loss binomial code. 
We only show results for the 1-loss binomial code because 
it outperforms the other RSBCs considered in this work.
At a total distance of $L_{tot}=1000$ km, $L_0=0.642$ km and decreases 
monotonically until at $L_{tot}=10000$ km,  $L_0=0.051$ km. 
We have chosen to fix the cost coefficient to a value similar to \cite{Muralidharan2014}, 
however we find the required elementary distances to be smaller in 
our case. In \cite{Muralidharan2014}, the authors found the optimal elementary
distances that minimize the cost coefficients. When we optimize 
the elementary distances, the cost coefficients will be reduced
significantly as shown in the inset of Fig. \ref{figcf} ($C^\prime_{opt}\sim10^{-4}$ at $L_{tot}=1000$ km).
However, in this case, the elementary distance $L_0$ is equal to $0.032$ km, thus requiring
a large number of stations. Although this is an impractical
scenario, if the number of stations available is large enough,
our scheme allows one to obtain much lower values of cost coefficients
than \cite{Muralidharan2014}. However, the realization of this 
scheme can be less demanding when higher-loss codes are in 
use. In fact, at loss order $l=7$, for a fixed value of $C^\prime=100$, $L_0$ can be greatly extended
up to $2.1$ km for a total distance $L_{tot}=1000$ km. In this case, $C^\prime_{opt}\sim10^{-5}$ at $L_0=0.7$ km, 
making it much more practical than the 1-loss code, and thus,
outperforming the third generation repeater scheme in \cite{Muralidharan2016}.

\section{\label{Con}Conclusion}
In this work, we have established the secret key rate for various rotation-symmetric bosonic codes, including the squeezed cat, binomial and 
GKP-like codes. 
Our results show that squeezing can be helpful to improve the overall performance of the repeater scheme (including 
SKR and increased elementary distance). 
The advantage appears for small squeezing ($r<0.05$) where the 
elementary distance $L_0$ in the range of $0.3$ km $\leq L_0 \leq 0.8$ km. Next the binomial and GKP-like codes outperform
cat codes significantly with the same loss order. 
Note that for practical use, the generation of squeezed cat codes may be more feasible
than binomial states with recent technologies. For instance, the generation of squeezed states was realized decades ago
using parametric down conversion \cite{Wu1986}, whereas the generation of binomial codes remains challenging, 
although the schemes to generate binomial states have been proposed theoretically \cite{LoFranco2010,Miry2014}.
As such, in the near term, squeezed cat codes are predicted to be more feasible for quantum communication. However,
the binomial codes could provide even better performance
as long as the generation of binomial states
is efficient.
Moreover, the binomial codes have a resource overhead comparable to,
and in some cases, less than, other third generation quantum repeaters.
However, the local losses will accumulate at every elementary 
station and inevitably decrease the SKR. Thus, further error-correcting 
elements such as additional quantum error correction codes on 
the atomic spin qubits need to be considered in future work. 
Overall the RSBCs are a good candidate for quantum repeater schemes in the near future because its requirements should be reachable with the state-of-the-art quantum devices.

\begin{acknowledgments}
    This work was supported by the Moonshot R\&D Program Grants JPMJMS2061 \& JPMJMS226C, the JSPS KAKENHI Grant No. 21H04880 and JST, the establishment of university fellowships towards the creation of science technology innovation, Grant Number JPMJFS2136.
\end{acknowledgments}

\section*{Conflict of Interest}
The authors declare no conflict of interest.

\appendix
\section{\label{flc}Full loss channel model}
Full channel loss can be modeled as follows \cite{Chuang1997},
\begin{eqnarray}
    \label{eqc8}
    \hat{\rho}'=\sum_{k=0}^{\infty}{\hat{A}_k\hat{\rho}\hat{A}_k^{\dagger}},
\end{eqnarray}
where $\hat{A}_k=\sqrt{(1-\eta)^k/k!}\sqrt{\eta}^{\hat{n}}\hat{a}^k$ is the 
Kraus operator that corresponds to the lost of $k$ photons, and $\eta$ is the transmission efficiency of the 
fiber. 
Hence the resulting joint atomic-photonic state after channel loss becomes 
\begin{eqnarray}
    \label{eqc9}
    \hat{\rho}_f=\sum_{k=0}^{\infty}{\hat{A}_k\ket{LM_{m,\Theta}}\bra{LM_{m,\Theta}}\hat{A}_k^{\dagger}},
\end{eqnarray}
where $\ket{LM_{m,\Theta}}=(\ket{\uparrow}\ket{0_{2^m,\Theta}}+\ket{\downarrow}\ket{1_{2^m,\Theta}})/\sqrt{2}$. Further after the syndrome measurement, the state will be projected to
\begin{eqnarray}
    \label{eqc10}
    \hat{\rho}_{sm}=\frac{1}{N_{sm}}\sum_{t=0}^{\infty}{\hat{A}_{j_t(m,q)}\ket{LM_{m,\Theta}}\bra{LM_{m,\Theta}}\hat{A}_{j_t(m,q)}^{\dagger}},
\end{eqnarray}
where $j_t(m,q)=2^m t+r_m(q)$ with 
$r_m(q)=q \mod 2^m$ and $N_{sm}$ is the normalization constant $N_{sm}={\rm tr}{\sum_{t=0}^{\infty}{\hat{A}_{j_t(m,q)}\ket{LM_{m,\Theta}}\bra{LM_{m,\Theta}}\hat{A}_{j_t(m,q)}^{\dagger}}}$.
After the entanglement creation process, we have
\\
\\
\\
\\
\begin{eqnarray}
    \label{eqc11}
    \begin{aligned}
        \hat{\rho}_{fi}=\frac{1}{N_{fi}}&\sum_{k=0}^{\infty} \hat{A}_{j_k(m,q)}\left(\ket{\phi_{j_k(m,q)}}\ket{0_{2^m,\Theta}}+\ket{\psi_{j_k(m,q)}}\ket{1_{2^m,\Theta}}\right)\\
        &\ \,\quad\times\left(\bra{\phi_{j_k(m,q)}}\bra{0_{2^m,\Theta}}+\bra{\psi_{j_k(m,q)}}\bra{1_{2^m,\Theta}}\right)\hat{A}_{j_k(m,q)}^{\dagger},
    \end{aligned}
\end{eqnarray}
where $N_{fi}$ is the normalization constant while $\ket{\phi_{j_k(m,q)}}$ and $\ket{\psi_{j_k(m,q)}}$ are defined as
\begin{eqnarray}
    \label{eqc15x}
    \begin{aligned}
        \ket{\phi_{j_k(m,q)}}&=\frac{\ket{\uparrow}\ket{\uparrow}+ e^{-\frac{ij_k(m,q)\pi}{2^m}}\ket{\downarrow}\ket{\downarrow}}{\sqrt{2}},\\
        \ket{\psi_{j_k(m,q)}}&=\frac{\ket{\uparrow}\ket{\downarrow}+ e^{-\frac{ij_k(m,q)\pi}{2^m}}\ket{\downarrow}\ket{\uparrow}}{\sqrt{2}}.
    \end{aligned}
\end{eqnarray}
In Eq. \ref{eqc11}, we have $j_k(m,q)/2^m=k+r_m(q)/2^m$, so the phase $\exp(ij_k(m,q)\pi/2^m)=(-1)^k\exp(-ir_m(q)\pi/2^m)$. 
Now $\hat{\rho}_{fi}$ can be then expressed as the sum of $\hat{\rho}_{fi,+}$ and $\hat{\rho}_{fi,-}$ defined by
\begin{widetext}    
    \begin{eqnarray}
        \label{eqc12}
        \begin{aligned}
            &\hat{\rho}_{fi,+}=\frac{1}{N_{fi,+}}\sum_{k\ even}^{\infty} \hat{A}_{j_{k}(m,q)}\left(\ket{\phi^+_{m,q}}\ket{0_{2^m,\Theta}}
            +\ket{\psi^+_{m,q}}\ket{1_{2^m,\Theta}}\right)
            \left(\bra{\phi^+_{m,q}}\bra{0_{2^m,\Theta}}
            +\bra{\psi^+_{m,q}}\bra{1_{2^m,\Theta}}\right)\hat{A}_{j_{k}(m,q)}^{\dagger},\\
            &\hat{\rho}_{fi,-}=\frac{1}{N_{fi,-}}\sum_{k\ odd}^{\infty} \hat{A}_{j_{k}(m,q)}\left(\ket{\phi^-_{m,q}}\ket{0_{2^m,\Theta}}
            +\ket{\psi^-_{m,q}}\ket{1_{2^m,\Theta}}\right)
            \left(\bra{\phi^-_{m,q}}\bra{0_{2^m,\Theta}}
            +\bra{\psi^-_{m,q}}\bra{1_{2^m,\Theta}}\right)\hat{A}_{j_{k}(m,q)}^{\dagger},
        \end{aligned}
    \end{eqnarray}
\end{widetext}  
where $N_{fi,+}$ and $N_{fi,-}$ are the normalization constants.
Here $\ket{\phi^\pm_{m,q}}$ and $\ket{\psi^\pm_{m,q}}$ are defined as 
\begin{eqnarray}
    \label{eqc15}
    \begin{aligned}
        \ket{\phi^\pm_{m,q}}&=\frac{\ket{\uparrow}\ket{\uparrow}\pm e^{-\frac{ir_m(q)\pi}{2^m}}\ket{\downarrow}\ket{\downarrow}}{\sqrt{2}},\\
        \ket{\psi^\pm_{m,q}}&=\frac{\ket{\uparrow}\ket{\downarrow}\pm e^{-\frac{ir_m(q)\pi}{2^m}}\ket{\downarrow}\ket{\uparrow}}{\sqrt{2}}.
    \end{aligned}
\end{eqnarray}

Next an approximation needs to be made in order to get the results after the
state discrimination process, which will be described in Appendix \ref{apx}. 
Following the discrimination between the states $\hat{A}_{r_m(q)}\ket{0_{2^m,\Theta}}$ 
and $\hat{A}_{r_m(q)}\ket{1_{2^m,\Theta}}$, the final atomic state will be either 
$\left(\ket{\uparrow}\ket{\uparrow}+e^{-\frac{iq\pi}{2^m}}\ket{\downarrow}\ket{\downarrow}\right)/\sqrt{2}$ or
$\left(\ket{\uparrow}\ket{\downarrow}-e^{-\frac{iq\pi}{2^m}}\ket{\downarrow}\ket{\uparrow}\right)/\sqrt{2}$ with some 
noisy terms
depending on the result of the state discrimination of the optical states. 
\section{\label{apx}Approximation for the SKR}
\begin{figure}
    \includegraphics[width=7.9cm]{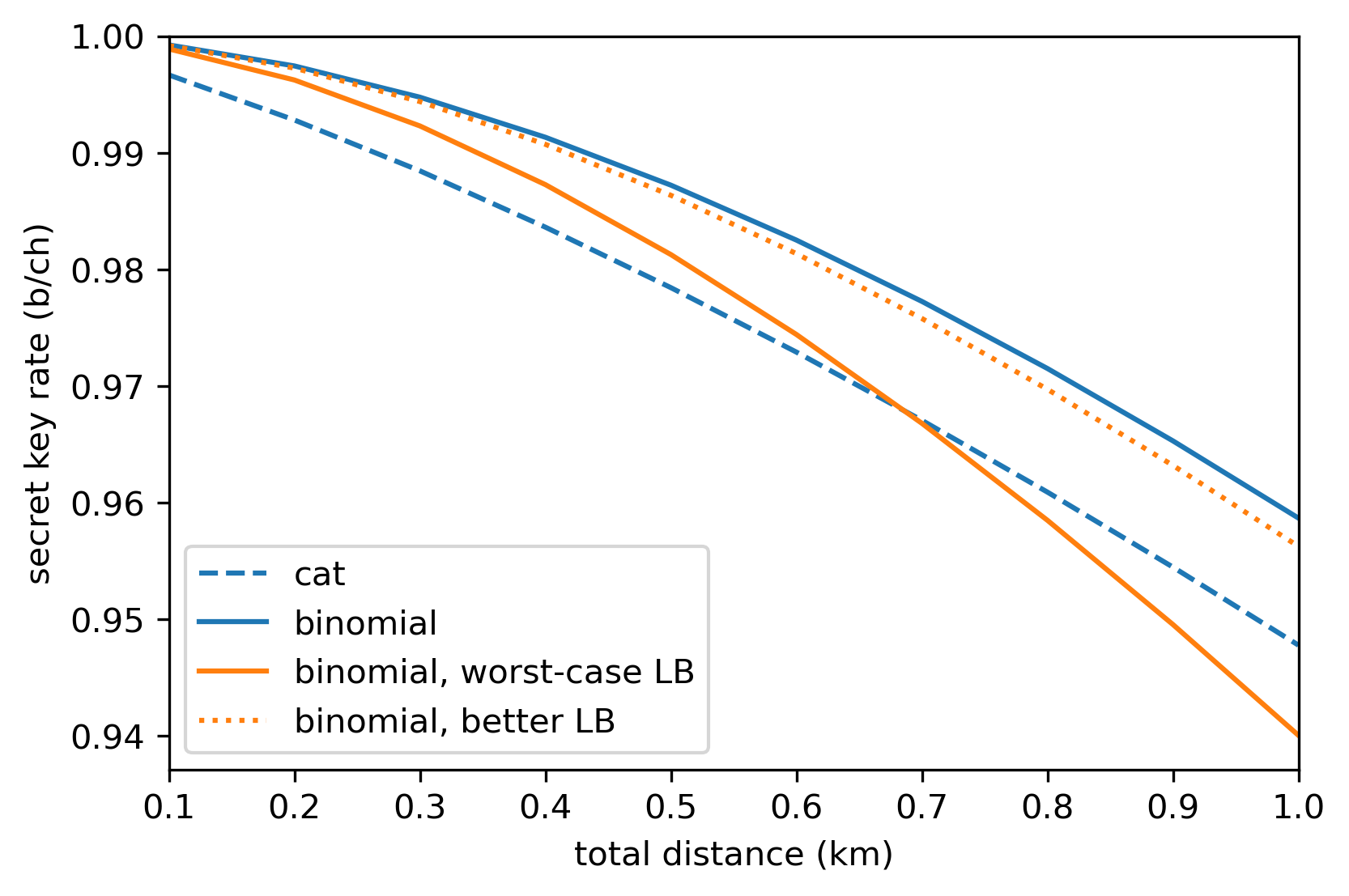}
    \captionsetup{justification=raggedright, singlelinecheck=false}
    \caption{\label{figab}The optimized SKR of the cat codes and the upper and lower bounds of 
    the optimized SKR of binomial codes with different 
    assumptions over the total 
    distance without any intermediate stations in between. The solid blue line represents the upper bound of 
    the SKR of the 
    binomial codes with the assumption that density matrix of losing $q+k(l+1)$ photons with $k\in \mathbb{N}$, $k\geq 1$
    is proportional to the 
    density matrix of losing $q$ photons. The orange solid line represents the worst-case lower bound of the SKR of the binomial codes.
    The dotted 
    orange line represents the better lower bound of the SKR of the binomial codes.
    }
\end{figure}
It is useful to point out that the codewords of the cat codes have the property:
\begin{eqnarray}
    \label{eqc13}
    \begin{aligned}
        \hat{A}_q\ket{C_{M,cat}}&=K_{M,q}\hat{A}_{q+M}\ket{C_{M,cat}},\\     
        \forall q&\in \mathbb{N}, \forall C=0,1,
    \end{aligned}
\end{eqnarray}
where $K_{M,q}$ is a constant that depends on $M$ and $q$. This means that the terms in Eq. \ref{eqc12} can 
be added up, but that is not always true for all types of RSBCs, e,g., squeezed cat codes, binomial codes or 
GKP-like codes. As for squeezed cat codes, the squeezing parameter $r$ considered in Fig. \ref{figsq} is 
very small so the codewords are not be so different from the corresponding 
cat codes (for $r=0$). We assume that the property described in Eq. \ref{eqc13} still holds for squeezed 
cat codes with small $r$. However, for general RSBCs other than cat codes, e.g., binomial codes, the property described in Eq. \ref{eqc13} does not hold anymore, so approximations have 
to be made before the numerical simulation. 
The approximation we make here is to treat all the undesired higher-loss terms as noisy terms and replace them
with the most noisy density matrix, the identity $\hat{I}$.
Here we replace the loss terms with $k\geq 1$ with identity $\hat{I}$, then the states after the entanglement creation 
process as shown in Eq. \ref{eqc11} will become
\begin{widetext}    
    \begin{eqnarray}
        \label{eqc14}
        \begin{aligned}
            \hat{\rho}_{fi}=&\ket{\phi^+_{m,q}}\bra{\phi^+_{m,q}}(\hat{A}_{j_0(m,q)}\ket{0_{2^m,\Theta}}\bra{0_{2^m,\Theta}}\hat{A}_{j_0(m,q)}^{\dagger}+w_0\hat{I})+\\
            &\ket{\psi^+_{m,q}}\bra{\psi^+_{m,q}}(\hat{A}_{j_0(m,q)}\ket{1_{2^m,\Theta}}\bra{1_{2^m,\Theta}}\hat{A}_{j_0(m,q)}^{\dagger}+w_0\hat{I})+\\
            &\ket{\phi^+_{m,q}}\bra{\phi^+_{m,q}}w_1\hat{I}+\ket{\psi^+_{m,q}}\bra{\psi^+_{m,q}}w_1\hat{I}+{\rm cross\ terms},
        \end{aligned}
    \end{eqnarray}
\end{widetext}
where $w_0$ and $w_1$ are the weights of the identity $\hat{I}$, which are equal to the norms of 
the sum of the density matrices with losing $q+k(l+1)$, i.e., $w_0={\rm tr} \sum_{k=1}^{\infty} \hat{A}_{j_{2k}(m,q)}\ket{0_{2^m,\Theta}}\bra{0_{2^m,\Theta}}\hat{A}_{j_{2k}(m,q)}^{\dagger}$ and 
$w_1={\rm tr} \sum_{k=1}^{\infty} \hat{A}_{j_{2k+1}(m,q)}\ket{0_{2^m,\Theta}}\bra{0_{2^m,\Theta}}\hat{A}_{j_{2k+1}(m,q)}^{\dagger}$. 
Then the results after the state
discrimination become straightforward since the identity $\hat{I}$ will always remain the same after any types of
measurements. The identity $\hat{I}$ does not contain any information and will increase the errors caused by 
the loss terms, thus the SKR we get with this approximation is smaller than its true value 
and thus it can be regarded as a lower bound of the SKR. We call it the worst-case lower bound of the 
fidelity of the binomial codes because the identity $\hat{I}$ introduces the most errors that this system can have.
The other lower bound of the fidelity is based on the overlap of the density matrix with losing $q+k(l+1)$ photons and the density matrix of losing $q$ photon. 
For the density matrix of losing $q$ photons, it can be written as 
\begin{widetext}
    \begin{eqnarray}
        \label{eqc16}
        \begin{aligned}
            &\hat{A}_q\ket{0_{2^m,\Theta}}\bra{0_{2^m,\Theta}}\hat{A}_q=v\hat{A}_q\ket{0_{2^m,\Theta}}\bra{0_{2^m,\Theta}}\hat{A}_q
            +(1-v)\hat{\rho}_{\perp},\\
        \end{aligned}
    \end{eqnarray}
\end{widetext}
where $v=\sqrt{F(\hat{A}_q\ket{0_{2^m,\Theta}}\bra{0_{2^m,\Theta}}\hat{A}_q,\hat{A}_0\ket{0_{2^m,\Theta}}\bra{0_{2^m,\Theta}}\hat{A}_0)}$ is 
the overlap of the density matrix of losing $q+k(l+1)$ photons $\hat{A}_q\ket{0_{2^m,\Theta}}\bra{0_{2^m,\Theta}}\hat{A}_q$ and the density matrix of losing $q$ photons $\hat{A}_0\ket{0_{2^m,\Theta}}\bra{0_{2^m,\Theta}}\hat{A}_0$ and $F(\hat{\rho},\hat{\sigma})=({\rm tr}\sqrt{\sqrt{\hat{\rho}}\hat{\sigma}\sqrt{\hat{\rho}}})^2$
is defined as the fidelity of any arbitrary density matrices $\hat{\rho}$ and $\hat{\sigma}$. $\hat{\rho}_{\perp}$ 
is the density matrix which is orthogonal to the 0-loss term $\hat{A}_0\ket{0_{2^m,\Theta}}\bra{0_{2^m,\Theta}}\hat{A}_0$,
i.e., $F(\hat{\rho}_{\perp},\hat{A}_0\ket{0_{2^m,\Theta}}\bra{0_{2^m,\Theta}}\hat{A}_0)=0$. Then the approximation 
is made by replacing $\hat{\rho}_{\perp}$ with identity $I$. With this approximation, we can obtain a better 
lower bound of the fidelity.

As shown in Fig. \ref{figab}, the worst-case lower bound of the SKR of the binomial code is a bit larger 
than the SKR of cat codes as the total distance is smaller than $0.86$ km, but it will be smaller than the 
SKR of cat code if total distance is larger than $0.86$ km. However, if we look at the better lower 
bound of the SKR of the binomial code, even it is a bit smaller than the upper bound of it, it is always larger
than the SKR of the cat code with the total distance larger than $0.1$ km and smaller than $1$ km. With the  
total distance larger than $0.1$ km and smaller than $1$ km, the overlap $v$ is roughly greater than $0.87$, which is not too far from
to $1$, meaning that the 2-loss term $\hat{A}_2\ket{0_{2^m,\Theta}}\bra{0_{2^m,\Theta}}\hat{A}_2$ is very close to the 
0-loss term $\hat{A}_0\ket{0_{2^m,\Theta}}\bra{0_{2^m,\Theta}}\hat{A}_0$ and they are almost proportional. 
Also, because the transmission coefficient $\eta$ with distance smaller than $1$ km is larger
than $0.955$, which is close to one, the probability to losing large number of photons is extremely low and 
the contribution of the density matrices with losing more than $l$ photons is very small.
These explain why this better lower bound is close to the upper bound with the assumption that density matrix of losing $q + k(l + 1)$ photons with
$k \in N, k \geq 1$ is proportional to the density matrix of
losing $q$ photons. Since $v$ is rather close to $1$ and from Fig. \ref{figab}, 
it is shown that the better lower bound and the upper bound is quite close to each other, we claim that 
the accurate value of the SKR is also close to the upper bound. Then it is reasonable to state that according to Fig. 
\ref{figbi}, the accurate value of SKR of the binomial code should also be much higher than the SKR of the 
cat code.

\bibliography{QP2}

\begin{thebibliography}{55}%
\makeatletter
\providecommand \@ifxundefined [1]{%
 \@ifx{#1\undefined}
}%
\providecommand \@ifnum [1]{%
 \ifnum #1\expandafter \@firstoftwo
 \else \expandafter \@secondoftwo
 \fi
}%
\providecommand \@ifx [1]{%
 \ifx #1\expandafter \@firstoftwo
 \else \expandafter \@secondoftwo
 \fi
}%
\providecommand \natexlab [1]{#1}%
\providecommand \enquote  [1]{``#1''}%
\providecommand \bibnamefont  [1]{#1}%
\providecommand \bibfnamefont [1]{#1}%
\providecommand \citenamefont [1]{#1}%
\providecommand \href@noop [0]{\@secondoftwo}%
\providecommand \href [0]{\begingroup \@sanitize@url \@href}%
\providecommand \@href[1]{\@@startlink{#1}\@@href}%
\providecommand \@@href[1]{\endgroup#1\@@endlink}%
\providecommand \@sanitize@url [0]{\catcode `\\12\catcode `\$12\catcode
  `\&12\catcode `\#12\catcode `\^12\catcode `\_12\catcode `\%12\relax}%
\providecommand \@@startlink[1]{}%
\providecommand \@@endlink[0]{}%
\providecommand \url  [0]{\begingroup\@sanitize@url \@url }%
\providecommand \@url [1]{\endgroup\@href {#1}{\urlprefix }}%
\providecommand \urlprefix  [0]{URL }%
\providecommand \Eprint [0]{\href }%
\providecommand \doibase [0]{https://doi.org/}%
\providecommand \selectlanguage [0]{\@gobble}%
\providecommand \bibinfo  [0]{\@secondoftwo}%
\providecommand \bibfield  [0]{\@secondoftwo}%
\providecommand \translation [1]{[#1]}%
\providecommand \BibitemOpen [0]{}%
\providecommand \bibitemStop [0]{}%
\providecommand \bibitemNoStop [0]{.\EOS\space}%
\providecommand \EOS [0]{\spacefactor3000\relax}%
\providecommand \BibitemShut  [1]{\csname bibitem#1\endcsname}%
\let\auto@bib@innerbib\@empty
\bibitem [{\citenamefont {Duan}\ \emph {et~al.}(2001)\citenamefont {Duan},
  \citenamefont {Lukin}, \citenamefont {Cirac},\ and\ \citenamefont
  {Zoller}}]{Duan2001}%
  \BibitemOpen
  \bibfield  {author} {\bibinfo {author} {\bibfnamefont {L.-M.}\ \bibnamefont
  {Duan}}, \bibinfo {author} {\bibfnamefont {M.~D.}\ \bibnamefont {Lukin}},
  \bibinfo {author} {\bibfnamefont {J.~I.}\ \bibnamefont {Cirac}},\ and\
  \bibinfo {author} {\bibfnamefont {P.}~\bibnamefont {Zoller}},\ }\bibfield
  {title} {\bibinfo {title} {Long-distance quantum communication with atomic
  ensembles and linear optics},\ }\href {https://doi.org/10.1038/35106500}
  {\bibfield  {journal} {\bibinfo  {journal} {Nature}\ }\textbf {\bibinfo
  {volume} {414}},\ \bibinfo {pages} {413} (\bibinfo {year}
  {2001})}\BibitemShut {NoStop}%
\bibitem [{\citenamefont {Jiang}\ \emph {et~al.}(2007)\citenamefont {Jiang},
  \citenamefont {Taylor},\ and\ \citenamefont {Lukin}}]{Jiang2007}%
  \BibitemOpen
  \bibfield  {author} {\bibinfo {author} {\bibfnamefont {L.}~\bibnamefont
  {Jiang}}, \bibinfo {author} {\bibfnamefont {J.~M.}\ \bibnamefont {Taylor}},\
  and\ \bibinfo {author} {\bibfnamefont {M.~D.}\ \bibnamefont {Lukin}},\
  }\bibfield  {title} {\bibinfo {title} {Fast and robust approach to
  long-distance quantum communication with atomic ensembles},\ }\href
  {https://doi.org/10.1103/PhysRevA.76.012301} {\bibfield  {journal} {\bibinfo
  {journal} {Phys. Rev. A}\ }\textbf {\bibinfo {volume} {76}},\ \bibinfo
  {pages} {012301} (\bibinfo {year} {2007})}\BibitemShut {NoStop}%
\bibitem [{\citenamefont {Muralidharan}\ \emph {et~al.}(2014)\citenamefont
  {Muralidharan}, \citenamefont {Kim}, \citenamefont {L\"utkenhaus},
  \citenamefont {Lukin},\ and\ \citenamefont {Jiang}}]{Muralidharan2014}%
  \BibitemOpen
  \bibfield  {author} {\bibinfo {author} {\bibfnamefont {S.}~\bibnamefont
  {Muralidharan}}, \bibinfo {author} {\bibfnamefont {J.}~\bibnamefont {Kim}},
  \bibinfo {author} {\bibfnamefont {N.}~\bibnamefont {L\"utkenhaus}}, \bibinfo
  {author} {\bibfnamefont {M.~D.}\ \bibnamefont {Lukin}},\ and\ \bibinfo
  {author} {\bibfnamefont {L.}~\bibnamefont {Jiang}},\ }\bibfield  {title}
  {\bibinfo {title} {Ultrafast and fault-tolerant quantum communication across
  long distances},\ }\href {https://doi.org/10.1103/PhysRevLett.112.250501}
  {\bibfield  {journal} {\bibinfo  {journal} {Phys. Rev. Lett.}\ }\textbf
  {\bibinfo {volume} {112}},\ \bibinfo {pages} {250501} (\bibinfo {year}
  {2014})}\BibitemShut {NoStop}%
\bibitem [{\citenamefont {Cirac}\ \emph {et~al.}(1997)\citenamefont {Cirac},
  \citenamefont {Zoller}, \citenamefont {Kimble},\ and\ \citenamefont
  {Mabuchi}}]{Cirac1997}%
  \BibitemOpen
  \bibfield  {author} {\bibinfo {author} {\bibfnamefont {J.~I.}\ \bibnamefont
  {Cirac}}, \bibinfo {author} {\bibfnamefont {P.}~\bibnamefont {Zoller}},
  \bibinfo {author} {\bibfnamefont {H.~J.}\ \bibnamefont {Kimble}},\ and\
  \bibinfo {author} {\bibfnamefont {H.}~\bibnamefont {Mabuchi}},\ }\bibfield
  {title} {\bibinfo {title} {Quantum state transfer and entanglement
  distribution among distant nodes in a quantum network},\ }\href
  {https://doi.org/10.1103/PhysRevLett.78.3221} {\bibfield  {journal} {\bibinfo
   {journal} {Phys. Rev. Lett.}\ }\textbf {\bibinfo {volume} {78}},\ \bibinfo
  {pages} {3221} (\bibinfo {year} {1997})}\BibitemShut {NoStop}%
\bibitem [{\citenamefont {Chiribella}\ \emph {et~al.}(2009)\citenamefont
  {Chiribella}, \citenamefont {D'Ariano},\ and\ \citenamefont
  {Perinotti}}]{Chiribella2009}%
  \BibitemOpen
  \bibfield  {author} {\bibinfo {author} {\bibfnamefont {G.}~\bibnamefont
  {Chiribella}}, \bibinfo {author} {\bibfnamefont {G.~M.}\ \bibnamefont
  {D'Ariano}},\ and\ \bibinfo {author} {\bibfnamefont {P.}~\bibnamefont
  {Perinotti}},\ }\bibfield  {title} {\bibinfo {title} {Theoretical framework
  for quantum networks},\ }\href {https://doi.org/10.1103/PhysRevA.80.022339}
  {\bibfield  {journal} {\bibinfo  {journal} {Phys. Rev. A}\ }\textbf {\bibinfo
  {volume} {80}},\ \bibinfo {pages} {022339} (\bibinfo {year}
  {2009})}\BibitemShut {NoStop}%
\bibitem [{\citenamefont {Perseguers}\ \emph {et~al.}(2010)\citenamefont
  {Perseguers}, \citenamefont {Lewenstein}, \citenamefont {Acín},\ and\
  \citenamefont {Cirac}}]{Perseguers2010}%
  \BibitemOpen
  \bibfield  {author} {\bibinfo {author} {\bibfnamefont {S.}~\bibnamefont
  {Perseguers}}, \bibinfo {author} {\bibfnamefont {M.}~\bibnamefont
  {Lewenstein}}, \bibinfo {author} {\bibfnamefont {A.}~\bibnamefont {Acín}},\
  and\ \bibinfo {author} {\bibfnamefont {J.~I.}\ \bibnamefont {Cirac}},\
  }\bibfield  {title} {\bibinfo {title} {Quantum random networks},\ }\href
  {https://doi.org/10.1038/nphys1665} {\bibfield  {journal} {\bibinfo
  {journal} {Nature Physics}\ }\textbf {\bibinfo {volume} {6}},\ \bibinfo
  {pages} {539} (\bibinfo {year} {2010})}\BibitemShut {NoStop}%
\bibitem [{\citenamefont {Kimble}(2008)}]{Kimble2008}%
  \BibitemOpen
  \bibfield  {author} {\bibinfo {author} {\bibfnamefont {H.~J.}\ \bibnamefont
  {Kimble}},\ }\bibfield  {title} {\bibinfo {title} {The quantum internet},\
  }\href {https://doi.org/10.1038/nature07127} {\bibfield  {journal} {\bibinfo
  {journal} {Nature}\ }\textbf {\bibinfo {volume} {453}},\ \bibinfo {pages}
  {1023} (\bibinfo {year} {2008})}\BibitemShut {NoStop}%
\bibitem [{\citenamefont {Wehner}\ \emph {et~al.}(2018)\citenamefont {Wehner},
  \citenamefont {Elkouss},\ and\ \citenamefont {Hanson}}]{Wehner2018}%
  \BibitemOpen
  \bibfield  {author} {\bibinfo {author} {\bibfnamefont {S.}~\bibnamefont
  {Wehner}}, \bibinfo {author} {\bibfnamefont {D.}~\bibnamefont {Elkouss}},\
  and\ \bibinfo {author} {\bibfnamefont {R.}~\bibnamefont {Hanson}},\
  }\bibfield  {title} {\bibinfo {title} {Quantum internet: A vision for the
  road ahead},\ }\href {https://doi.org/10.1126/science.aam9288} {\bibfield
  {journal} {\bibinfo  {journal} {Science}\ }\textbf {\bibinfo {volume}
  {362}},\ \bibinfo {pages} {eaam9288} (\bibinfo {year} {2018})}\BibitemShut
  {NoStop}%
\bibitem [{\citenamefont {Munro}\ \emph {et~al.}(2022)\citenamefont {Munro},
  \citenamefont {Piparo}, \citenamefont {Dias}, \citenamefont {Hanks},\ and\
  \citenamefont {Nemoto}}]{Munro2022}%
  \BibitemOpen
  \bibfield  {author} {\bibinfo {author} {\bibfnamefont {W.~J.}\ \bibnamefont
  {Munro}}, \bibinfo {author} {\bibfnamefont {N.~L.}\ \bibnamefont {Piparo}},
  \bibinfo {author} {\bibfnamefont {J.}~\bibnamefont {Dias}}, \bibinfo {author}
  {\bibfnamefont {M.}~\bibnamefont {Hanks}},\ and\ \bibinfo {author}
  {\bibfnamefont {K.}~\bibnamefont {Nemoto}},\ }\bibfield  {title} {\bibinfo
  {title} {Designing tomorrow{\textquotesingle}s quantum internet},\ }\href
  {https://doi.org/10.1116/5.0092069} {\bibfield  {journal} {\bibinfo
  {journal} {{AVS} Quantum Science}\ }\textbf {\bibinfo {volume} {4}},\
  \bibinfo {pages} {020503} (\bibinfo {year} {2022})}\BibitemShut {NoStop}%
\bibitem [{\citenamefont {O'Brien}(2007)}]{OBrien2007}%
  \BibitemOpen
  \bibfield  {author} {\bibinfo {author} {\bibfnamefont {J.~L.}\ \bibnamefont
  {O'Brien}},\ }\bibfield  {title} {\bibinfo {title} {Optical quantum
  computing},\ }\href {https://doi.org/10.1126/science.1142892} {\bibfield
  {journal} {\bibinfo  {journal} {Science}\ }\textbf {\bibinfo {volume}
  {318}},\ \bibinfo {pages} {1567} (\bibinfo {year} {2007})}\BibitemShut
  {NoStop}%
\bibitem [{\citenamefont {Bandyopadhyay}(2000)}]{Bandyopadhyay2000}%
  \BibitemOpen
  \bibfield  {author} {\bibinfo {author} {\bibfnamefont {S.}~\bibnamefont
  {Bandyopadhyay}},\ }\bibfield  {title} {\bibinfo {title} {Teleportation and
  secret sharing with pure entangled states},\ }\href
  {https://doi.org/10.1103/PhysRevA.62.012308} {\bibfield  {journal} {\bibinfo
  {journal} {Phys. Rev. A}\ }\textbf {\bibinfo {volume} {62}},\ \bibinfo
  {pages} {012308} (\bibinfo {year} {2000})}\BibitemShut {NoStop}%
\bibitem [{\citenamefont {Ikram}\ \emph {et~al.}(2000)\citenamefont {Ikram},
  \citenamefont {Zhu},\ and\ \citenamefont {Zubairy}}]{Ikram2000}%
  \BibitemOpen
  \bibfield  {author} {\bibinfo {author} {\bibfnamefont {M.}~\bibnamefont
  {Ikram}}, \bibinfo {author} {\bibfnamefont {S.-Y.}\ \bibnamefont {Zhu}},\
  and\ \bibinfo {author} {\bibfnamefont {M.~S.}\ \bibnamefont {Zubairy}},\
  }\bibfield  {title} {\bibinfo {title} {Quantum teleportation of an entangled
  state},\ }\href {https://doi.org/10.1103/PhysRevA.62.022307} {\bibfield
  {journal} {\bibinfo  {journal} {Phys. Rev. A}\ }\textbf {\bibinfo {volume}
  {62}},\ \bibinfo {pages} {022307} (\bibinfo {year} {2000})}\BibitemShut
  {NoStop}%
\bibitem [{\citenamefont {Shi}\ \emph {et~al.}(2001)\citenamefont {Shi},
  \citenamefont {Li}, \citenamefont {Liu}, \citenamefont {Fan},\ and\
  \citenamefont {Guo}}]{Shi2001}%
  \BibitemOpen
  \bibfield  {author} {\bibinfo {author} {\bibfnamefont {B.-S.}\ \bibnamefont
  {Shi}}, \bibinfo {author} {\bibfnamefont {J.}~\bibnamefont {Li}}, \bibinfo
  {author} {\bibfnamefont {J.-M.}\ \bibnamefont {Liu}}, \bibinfo {author}
  {\bibfnamefont {X.-F.}\ \bibnamefont {Fan}},\ and\ \bibinfo {author}
  {\bibfnamefont {G.-C.}\ \bibnamefont {Guo}},\ }\bibfield  {title} {\bibinfo
  {title} {Quantum key distribution and quantum authentication based on
  entangled state},\ }\href
  {https://doi.org/https://doi.org/10.1016/S0375-9601(01)00129-3} {\bibfield
  {journal} {\bibinfo  {journal} {Physics Letters A}\ }\textbf {\bibinfo
  {volume} {281}},\ \bibinfo {pages} {83} (\bibinfo {year} {2001})}\BibitemShut
  {NoStop}%
\bibitem [{\citenamefont {Briegel}\ \emph {et~al.}(1998)\citenamefont
  {Briegel}, \citenamefont {D\"ur}, \citenamefont {Cirac},\ and\ \citenamefont
  {Zoller}}]{Briegel1998}%
  \BibitemOpen
  \bibfield  {author} {\bibinfo {author} {\bibfnamefont {H.-J.}\ \bibnamefont
  {Briegel}}, \bibinfo {author} {\bibfnamefont {W.}~\bibnamefont {D\"ur}},
  \bibinfo {author} {\bibfnamefont {J.~I.}\ \bibnamefont {Cirac}},\ and\
  \bibinfo {author} {\bibfnamefont {P.}~\bibnamefont {Zoller}},\ }\bibfield
  {title} {\bibinfo {title} {Quantum repeaters: The role of imperfect local
  operations in quantum communication},\ }\href
  {https://doi.org/10.1103/PhysRevLett.81.5932} {\bibfield  {journal} {\bibinfo
   {journal} {Phys. Rev. Lett.}\ }\textbf {\bibinfo {volume} {81}},\ \bibinfo
  {pages} {5932} (\bibinfo {year} {1998})}\BibitemShut {NoStop}%
\bibitem [{\citenamefont {D\"ur}\ \emph {et~al.}(1999)\citenamefont {D\"ur},
  \citenamefont {Briegel}, \citenamefont {Cirac},\ and\ \citenamefont
  {Zoller}}]{Duer1999}%
  \BibitemOpen
  \bibfield  {author} {\bibinfo {author} {\bibfnamefont {W.}~\bibnamefont
  {D\"ur}}, \bibinfo {author} {\bibfnamefont {H.-J.}\ \bibnamefont {Briegel}},
  \bibinfo {author} {\bibfnamefont {J.~I.}\ \bibnamefont {Cirac}},\ and\
  \bibinfo {author} {\bibfnamefont {P.}~\bibnamefont {Zoller}},\ }\bibfield
  {title} {\bibinfo {title} {Quantum repeaters based on entanglement
  purification},\ }\href {https://doi.org/10.1103/PhysRevA.59.169} {\bibfield
  {journal} {\bibinfo  {journal} {Phys. Rev. A}\ }\textbf {\bibinfo {volume}
  {59}},\ \bibinfo {pages} {169} (\bibinfo {year} {1999})}\BibitemShut
  {NoStop}%
\bibitem [{\citenamefont {van Loock}\ \emph {et~al.}(2006)\citenamefont {van
  Loock}, \citenamefont {Ladd}, \citenamefont {Sanaka}, \citenamefont
  {Yamaguchi}, \citenamefont {Nemoto}, \citenamefont {Munro},\ and\
  \citenamefont {Yamamoto}}]{Loock2006}%
  \BibitemOpen
  \bibfield  {author} {\bibinfo {author} {\bibfnamefont {P.}~\bibnamefont {van
  Loock}}, \bibinfo {author} {\bibfnamefont {T.~D.}\ \bibnamefont {Ladd}},
  \bibinfo {author} {\bibfnamefont {K.}~\bibnamefont {Sanaka}}, \bibinfo
  {author} {\bibfnamefont {F.}~\bibnamefont {Yamaguchi}}, \bibinfo {author}
  {\bibfnamefont {K.}~\bibnamefont {Nemoto}}, \bibinfo {author} {\bibfnamefont
  {W.~J.}\ \bibnamefont {Munro}},\ and\ \bibinfo {author} {\bibfnamefont
  {Y.}~\bibnamefont {Yamamoto}},\ }\bibfield  {title} {\bibinfo {title} {Hybrid
  quantum repeater using bright coherent light},\ }\href
  {https://doi.org/10.1103/PhysRevLett.96.240501} {\bibfield  {journal}
  {\bibinfo  {journal} {Phys. Rev. Lett.}\ }\textbf {\bibinfo {volume} {96}},\
  \bibinfo {pages} {240501} (\bibinfo {year} {2006})}\BibitemShut {NoStop}%
\bibitem [{\citenamefont {Jiang}\ \emph {et~al.}(2009)\citenamefont {Jiang},
  \citenamefont {Taylor}, \citenamefont {Nemoto}, \citenamefont {Munro},
  \citenamefont {Van~Meter},\ and\ \citenamefont {Lukin}}]{Jiang2009}%
  \BibitemOpen
  \bibfield  {author} {\bibinfo {author} {\bibfnamefont {L.}~\bibnamefont
  {Jiang}}, \bibinfo {author} {\bibfnamefont {J.~M.}\ \bibnamefont {Taylor}},
  \bibinfo {author} {\bibfnamefont {K.}~\bibnamefont {Nemoto}}, \bibinfo
  {author} {\bibfnamefont {W.~J.}\ \bibnamefont {Munro}}, \bibinfo {author}
  {\bibfnamefont {R.}~\bibnamefont {Van~Meter}},\ and\ \bibinfo {author}
  {\bibfnamefont {M.~D.}\ \bibnamefont {Lukin}},\ }\bibfield  {title} {\bibinfo
  {title} {Quantum repeater with encoding},\ }\href
  {https://doi.org/10.1103/PhysRevA.79.032325} {\bibfield  {journal} {\bibinfo
  {journal} {Phys. Rev. A}\ }\textbf {\bibinfo {volume} {79}},\ \bibinfo
  {pages} {032325} (\bibinfo {year} {2009})}\BibitemShut {NoStop}%
\bibitem [{\citenamefont {Sangouard}\ \emph {et~al.}(2011)\citenamefont
  {Sangouard}, \citenamefont {Simon}, \citenamefont {de~Riedmatten},\ and\
  \citenamefont {Gisin}}]{Sangouard2011}%
  \BibitemOpen
  \bibfield  {author} {\bibinfo {author} {\bibfnamefont {N.}~\bibnamefont
  {Sangouard}}, \bibinfo {author} {\bibfnamefont {C.}~\bibnamefont {Simon}},
  \bibinfo {author} {\bibfnamefont {H.}~\bibnamefont {de~Riedmatten}},\ and\
  \bibinfo {author} {\bibfnamefont {N.}~\bibnamefont {Gisin}},\ }\bibfield
  {title} {\bibinfo {title} {Quantum repeaters based on atomic ensembles and
  linear optics},\ }\href {https://doi.org/10.1103/RevModPhys.83.33} {\bibfield
   {journal} {\bibinfo  {journal} {Rev. Mod. Phys.}\ }\textbf {\bibinfo
  {volume} {83}},\ \bibinfo {pages} {33} (\bibinfo {year} {2011})}\BibitemShut
  {NoStop}%
\bibitem [{\citenamefont {Munro}\ \emph {et~al.}(2015)\citenamefont {Munro},
  \citenamefont {Azuma}, \citenamefont {Tamaki},\ and\ \citenamefont
  {Nemoto}}]{Munro2015}%
  \BibitemOpen
  \bibfield  {author} {\bibinfo {author} {\bibfnamefont {W.~J.}\ \bibnamefont
  {Munro}}, \bibinfo {author} {\bibfnamefont {K.}~\bibnamefont {Azuma}},
  \bibinfo {author} {\bibfnamefont {K.}~\bibnamefont {Tamaki}},\ and\ \bibinfo
  {author} {\bibfnamefont {K.}~\bibnamefont {Nemoto}},\ }\bibfield  {title}
  {\bibinfo {title} {Inside quantum repeaters},\ }\href
  {https://doi.org/10.1109/jstqe.2015.2392076} {\bibfield  {journal} {\bibinfo
  {journal} {{IEEE} Journal of Selected Topics in Quantum Electronics}\
  }\textbf {\bibinfo {volume} {21}},\ \bibinfo {pages} {78} (\bibinfo {year}
  {2015})}\BibitemShut {NoStop}%
\bibitem [{\citenamefont {Azuma}\ \emph {et~al.}(2015)\citenamefont {Azuma},
  \citenamefont {Tamaki},\ and\ \citenamefont {Lo}}]{Azuma2015}%
  \BibitemOpen
  \bibfield  {author} {\bibinfo {author} {\bibfnamefont {K.}~\bibnamefont
  {Azuma}}, \bibinfo {author} {\bibfnamefont {K.}~\bibnamefont {Tamaki}},\ and\
  \bibinfo {author} {\bibfnamefont {H.-K.}\ \bibnamefont {Lo}},\ }\bibfield
  {title} {\bibinfo {title} {All-photonic quantum repeaters},\ }\href
  {https://doi.org/10.1038/ncomms7787} {\bibfield  {journal} {\bibinfo
  {journal} {Nature Communications}\ }\textbf {\bibinfo {volume} {6}},\
  \bibinfo {pages} {6787} (\bibinfo {year} {2015})}\BibitemShut {NoStop}%
\bibitem [{\citenamefont {Nemoto}\ \emph {et~al.}(2016)\citenamefont {Nemoto},
  \citenamefont {Trupke}, \citenamefont {Devitt}, \citenamefont
  {Scharfenberger}, \citenamefont {Buczak}, \citenamefont {Schmiedmayer},\ and\
  \citenamefont {Munro}}]{Nemoto2016}%
  \BibitemOpen
  \bibfield  {author} {\bibinfo {author} {\bibfnamefont {K.}~\bibnamefont
  {Nemoto}}, \bibinfo {author} {\bibfnamefont {M.}~\bibnamefont {Trupke}},
  \bibinfo {author} {\bibfnamefont {S.~J.}\ \bibnamefont {Devitt}}, \bibinfo
  {author} {\bibfnamefont {B.}~\bibnamefont {Scharfenberger}}, \bibinfo
  {author} {\bibfnamefont {K.}~\bibnamefont {Buczak}}, \bibinfo {author}
  {\bibfnamefont {J.}~\bibnamefont {Schmiedmayer}},\ and\ \bibinfo {author}
  {\bibfnamefont {W.~J.}\ \bibnamefont {Munro}},\ }\bibfield  {title} {\bibinfo
  {title} {Photonic quantum networks formed from nv- centers},\ }\href
  {https://doi.org/10.1038/srep26284} {\bibfield  {journal} {\bibinfo
  {journal} {Scientific Reports}\ }\textbf {\bibinfo {volume} {6}},\ \bibinfo
  {pages} {26284} (\bibinfo {year} {2016})}\BibitemShut {NoStop}%
\bibitem [{\citenamefont {Bennett}\ \emph {et~al.}(1996)\citenamefont
  {Bennett}, \citenamefont {Brassard}, \citenamefont {Popescu}, \citenamefont
  {Schumacher}, \citenamefont {Smolin},\ and\ \citenamefont
  {Wootters}}]{Bennett1996}%
  \BibitemOpen
  \bibfield  {author} {\bibinfo {author} {\bibfnamefont {C.~H.}\ \bibnamefont
  {Bennett}}, \bibinfo {author} {\bibfnamefont {G.}~\bibnamefont {Brassard}},
  \bibinfo {author} {\bibfnamefont {S.}~\bibnamefont {Popescu}}, \bibinfo
  {author} {\bibfnamefont {B.}~\bibnamefont {Schumacher}}, \bibinfo {author}
  {\bibfnamefont {J.~A.}\ \bibnamefont {Smolin}},\ and\ \bibinfo {author}
  {\bibfnamefont {W.~K.}\ \bibnamefont {Wootters}},\ }\bibfield  {title}
  {\bibinfo {title} {Purification of noisy entanglement and faithful
  teleportation via noisy channels},\ }\href
  {https://doi.org/10.1103/PhysRevLett.76.722} {\bibfield  {journal} {\bibinfo
  {journal} {Phys. Rev. Lett.}\ }\textbf {\bibinfo {volume} {76}},\ \bibinfo
  {pages} {722} (\bibinfo {year} {1996})}\BibitemShut {NoStop}%
\bibitem [{\citenamefont {\ifmmode~\dot{Z}\else \.{Z}\fi{}ukowski}\ \emph
  {et~al.}(1993)\citenamefont {\ifmmode~\dot{Z}\else \.{Z}\fi{}ukowski},
  \citenamefont {Zeilinger}, \citenamefont {Horne},\ and\ \citenamefont
  {Ekert}}]{ifmmodeZelseZfiukowski1993}%
  \BibitemOpen
  \bibfield  {author} {\bibinfo {author} {\bibfnamefont {M.}~\bibnamefont
  {\ifmmode~\dot{Z}\else \.{Z}\fi{}ukowski}}, \bibinfo {author} {\bibfnamefont
  {A.}~\bibnamefont {Zeilinger}}, \bibinfo {author} {\bibfnamefont {M.~A.}\
  \bibnamefont {Horne}},\ and\ \bibinfo {author} {\bibfnamefont {A.~K.}\
  \bibnamefont {Ekert}},\ }\bibfield  {title} {\bibinfo {title}
  {``event-ready-detectors'' bell experiment via entanglement swapping},\
  }\href {https://doi.org/10.1103/PhysRevLett.71.4287} {\bibfield  {journal}
  {\bibinfo  {journal} {Phys. Rev. Lett.}\ }\textbf {\bibinfo {volume} {71}},\
  \bibinfo {pages} {4287} (\bibinfo {year} {1993})}\BibitemShut {NoStop}%
\bibitem [{\citenamefont {Pan}\ \emph {et~al.}(2001)\citenamefont {Pan},
  \citenamefont {Daniell}, \citenamefont {Gasparoni}, \citenamefont {Weihs},\
  and\ \citenamefont {Zeilinger}}]{Pan2001}%
  \BibitemOpen
  \bibfield  {author} {\bibinfo {author} {\bibfnamefont {J.-W.}\ \bibnamefont
  {Pan}}, \bibinfo {author} {\bibfnamefont {M.}~\bibnamefont {Daniell}},
  \bibinfo {author} {\bibfnamefont {S.}~\bibnamefont {Gasparoni}}, \bibinfo
  {author} {\bibfnamefont {G.}~\bibnamefont {Weihs}},\ and\ \bibinfo {author}
  {\bibfnamefont {A.}~\bibnamefont {Zeilinger}},\ }\bibfield  {title} {\bibinfo
  {title} {Experimental demonstration of four-photon entanglement and
  high-fidelity teleportation},\ }\href
  {https://doi.org/10.1103/PhysRevLett.86.4435} {\bibfield  {journal} {\bibinfo
   {journal} {Phys. Rev. Lett.}\ }\textbf {\bibinfo {volume} {86}},\ \bibinfo
  {pages} {4435} (\bibinfo {year} {2001})}\BibitemShut {NoStop}%
\bibitem [{\citenamefont {Munro}\ \emph {et~al.}(2010)\citenamefont {Munro},
  \citenamefont {Harrison}, \citenamefont {Stephens}, \citenamefont {Devitt},\
  and\ \citenamefont {Nemoto}}]{Munro2010}%
  \BibitemOpen
  \bibfield  {author} {\bibinfo {author} {\bibfnamefont {W.~J.}\ \bibnamefont
  {Munro}}, \bibinfo {author} {\bibfnamefont {K.~A.}\ \bibnamefont {Harrison}},
  \bibinfo {author} {\bibfnamefont {A.~M.}\ \bibnamefont {Stephens}}, \bibinfo
  {author} {\bibfnamefont {S.~J.}\ \bibnamefont {Devitt}},\ and\ \bibinfo
  {author} {\bibfnamefont {K.}~\bibnamefont {Nemoto}},\ }\bibfield  {title}
  {\bibinfo {title} {From quantum multiplexing to high-performance quantum
  networking},\ }\href {https://doi.org/10.1038/nphoton.2010.213} {\bibfield
  {journal} {\bibinfo  {journal} {Nature Photonics}\ }\textbf {\bibinfo
  {volume} {4}},\ \bibinfo {pages} {792} (\bibinfo {year} {2010})}\BibitemShut
  {NoStop}%
\bibitem [{\citenamefont {Zwerger}\ \emph {et~al.}(2014)\citenamefont
  {Zwerger}, \citenamefont {Briegel},\ and\ \citenamefont
  {Dür}}]{Zwerger2014}%
  \BibitemOpen
  \bibfield  {author} {\bibinfo {author} {\bibfnamefont {M.}~\bibnamefont
  {Zwerger}}, \bibinfo {author} {\bibfnamefont {H.~J.}\ \bibnamefont
  {Briegel}},\ and\ \bibinfo {author} {\bibfnamefont {W.}~\bibnamefont
  {Dür}},\ }\bibfield  {title} {\bibinfo {title} {Hybrid architecture for
  encoded measurement-based quantum computation},\ }\href
  {https://doi.org/10.1038/srep05364} {\bibfield  {journal} {\bibinfo
  {journal} {Scientific Reports}\ }\textbf {\bibinfo {volume} {4}},\ \bibinfo
  {pages} {5364} (\bibinfo {year} {2014})}\BibitemShut {NoStop}%
\bibitem [{\citenamefont {Fowler}\ \emph {et~al.}(2010)\citenamefont {Fowler},
  \citenamefont {Wang}, \citenamefont {Hill}, \citenamefont {Ladd},
  \citenamefont {Van~Meter},\ and\ \citenamefont {Hollenberg}}]{Fowler2010}%
  \BibitemOpen
  \bibfield  {author} {\bibinfo {author} {\bibfnamefont {A.~G.}\ \bibnamefont
  {Fowler}}, \bibinfo {author} {\bibfnamefont {D.~S.}\ \bibnamefont {Wang}},
  \bibinfo {author} {\bibfnamefont {C.~D.}\ \bibnamefont {Hill}}, \bibinfo
  {author} {\bibfnamefont {T.~D.}\ \bibnamefont {Ladd}}, \bibinfo {author}
  {\bibfnamefont {R.}~\bibnamefont {Van~Meter}},\ and\ \bibinfo {author}
  {\bibfnamefont {L.~C.~L.}\ \bibnamefont {Hollenberg}},\ }\bibfield  {title}
  {\bibinfo {title} {Surface code quantum communication},\ }\href
  {https://doi.org/10.1103/PhysRevLett.104.180503} {\bibfield  {journal}
  {\bibinfo  {journal} {Phys. Rev. Lett.}\ }\textbf {\bibinfo {volume} {104}},\
  \bibinfo {pages} {180503} (\bibinfo {year} {2010})}\BibitemShut {NoStop}%
\bibitem [{\citenamefont {Munro}\ \emph {et~al.}(2012)\citenamefont {Munro},
  \citenamefont {Stephens}, \citenamefont {Devitt}, \citenamefont {Harrison},\
  and\ \citenamefont {Nemoto}}]{Munro2012}%
  \BibitemOpen
  \bibfield  {author} {\bibinfo {author} {\bibfnamefont {W.~J.}\ \bibnamefont
  {Munro}}, \bibinfo {author} {\bibfnamefont {A.~M.}\ \bibnamefont {Stephens}},
  \bibinfo {author} {\bibfnamefont {S.~J.}\ \bibnamefont {Devitt}}, \bibinfo
  {author} {\bibfnamefont {K.~A.}\ \bibnamefont {Harrison}},\ and\ \bibinfo
  {author} {\bibfnamefont {K.}~\bibnamefont {Nemoto}},\ }\bibfield  {title}
  {\bibinfo {title} {Quantum communication without the necessity of quantum
  memories},\ }\href {https://doi.org/10.1038/nphoton.2012.243} {\bibfield
  {journal} {\bibinfo  {journal} {Nature Photonics}\ }\textbf {\bibinfo
  {volume} {6}},\ \bibinfo {pages} {777} (\bibinfo {year} {2012})}\BibitemShut
  {NoStop}%
\bibitem [{\citenamefont {Muralidharan}\ \emph {et~al.}(2016)\citenamefont
  {Muralidharan}, \citenamefont {Li}, \citenamefont {Kim}, \citenamefont
  {Lütkenhaus}, \citenamefont {Lukin},\ and\ \citenamefont
  {Jiang}}]{Muralidharan2016}%
  \BibitemOpen
  \bibfield  {author} {\bibinfo {author} {\bibfnamefont {S.}~\bibnamefont
  {Muralidharan}}, \bibinfo {author} {\bibfnamefont {L.}~\bibnamefont {Li}},
  \bibinfo {author} {\bibfnamefont {J.}~\bibnamefont {Kim}}, \bibinfo {author}
  {\bibfnamefont {N.}~\bibnamefont {Lütkenhaus}}, \bibinfo {author}
  {\bibfnamefont {M.~D.}\ \bibnamefont {Lukin}},\ and\ \bibinfo {author}
  {\bibfnamefont {L.}~\bibnamefont {Jiang}},\ }\bibfield  {title} {\bibinfo
  {title} {Optimal architectures for long distance quantum communication},\
  }\href {https://doi.org/10.1038/srep20463} {\bibfield  {journal} {\bibinfo
  {journal} {Scientific Reports}\ }\textbf {\bibinfo {volume} {6}},\ \bibinfo
  {pages} {20463} (\bibinfo {year} {2016})}\BibitemShut {NoStop}%
\bibitem [{\citenamefont {Li}\ and\ \citenamefont {van Loock}(2023)}]{Li2023}%
  \BibitemOpen
  \bibfield  {author} {\bibinfo {author} {\bibfnamefont {P.-Z.}\ \bibnamefont
  {Li}}\ and\ \bibinfo {author} {\bibfnamefont {P.}~\bibnamefont {van Loock}},\
  }\bibfield  {title} {\bibinfo {title} {Memoryless quantum repeaters based on
  cavity-qed and coherent states},\ }\href
  {https://doi.org/10.1002/qute.202200151} {\bibfield  {journal} {\bibinfo
  {journal} {Adv Quantum Technol.}\ ,\ \bibinfo {pages} {2200151}} (\bibinfo
  {year} {2023})}\BibitemShut {NoStop}%
\bibitem [{\citenamefont {Grimsmo}\ \emph {et~al.}(2020)\citenamefont
  {Grimsmo}, \citenamefont {Combes},\ and\ \citenamefont
  {Baragiola}}]{Grimsmo2020}%
  \BibitemOpen
  \bibfield  {author} {\bibinfo {author} {\bibfnamefont {A.~L.}\ \bibnamefont
  {Grimsmo}}, \bibinfo {author} {\bibfnamefont {J.}~\bibnamefont {Combes}},\
  and\ \bibinfo {author} {\bibfnamefont {B.~Q.}\ \bibnamefont {Baragiola}},\
  }\bibfield  {title} {\bibinfo {title} {Quantum computing with
  rotation-symmetric bosonic codes},\ }\href
  {https://doi.org/10.1103/PhysRevX.10.011058} {\bibfield  {journal} {\bibinfo
  {journal} {Physical Review X}\ }\textbf {\bibinfo {volume} {10}},\ \bibinfo
  {pages} {011058} (\bibinfo {year} {2020})}\BibitemShut {NoStop}%
\bibitem [{\citenamefont {Liu}\ \emph {et~al.}(2005)\citenamefont {Liu},
  \citenamefont {Wei},\ and\ \citenamefont {Nori}}]{Liu2005}%
  \BibitemOpen
  \bibfield  {author} {\bibinfo {author} {\bibfnamefont {Y.-x.}\ \bibnamefont
  {Liu}}, \bibinfo {author} {\bibfnamefont {L.~F.}\ \bibnamefont {Wei}},\ and\
  \bibinfo {author} {\bibfnamefont {F.}~\bibnamefont {Nori}},\ }\bibfield
  {title} {\bibinfo {title} {Preparation of macroscopic quantum superposition
  states of a cavity field via coupling to a superconducting charge qubit},\
  }\href {https://doi.org/10.1103/PhysRevA.71.063820} {\bibfield  {journal}
  {\bibinfo  {journal} {Phys. Rev. A}\ }\textbf {\bibinfo {volume} {71}},\
  \bibinfo {pages} {063820} (\bibinfo {year} {2005})}\BibitemShut {NoStop}%
\bibitem [{\citenamefont {Schlegel}\ \emph {et~al.}(2022)\citenamefont
  {Schlegel}, \citenamefont {Minganti},\ and\ \citenamefont
  {Savona}}]{Schlegel2022}%
  \BibitemOpen
  \bibfield  {author} {\bibinfo {author} {\bibfnamefont {D.~S.}\ \bibnamefont
  {Schlegel}}, \bibinfo {author} {\bibfnamefont {F.}~\bibnamefont {Minganti}},\
  and\ \bibinfo {author} {\bibfnamefont {V.}~\bibnamefont {Savona}},\
  }\bibfield  {title} {\bibinfo {title} {Quantum error correction using
  squeezed schrödinger cat states},\ }\href
  {https://doi.org/10.1103/physreva.106.022431} {\bibfield  {journal} {\bibinfo
   {journal} {Physical Review A}\ }\textbf {\bibinfo {volume} {106}},\ \bibinfo
  {pages} {022431} (\bibinfo {year} {2022})}\BibitemShut {NoStop}%
\bibitem [{\citenamefont {Xu}\ \emph {et~al.}(2022{\natexlab{a}})\citenamefont
  {Xu}, \citenamefont {Zheng}, \citenamefont {Wang}, \citenamefont {Zoller},
  \citenamefont {Clerk},\ and\ \citenamefont {Jiang}}]{xu2022b}%
  \BibitemOpen
  \bibfield  {author} {\bibinfo {author} {\bibfnamefont {Q.}~\bibnamefont
  {Xu}}, \bibinfo {author} {\bibfnamefont {G.}~\bibnamefont {Zheng}}, \bibinfo
  {author} {\bibfnamefont {Y.-X.}\ \bibnamefont {Wang}}, \bibinfo {author}
  {\bibfnamefont {P.}~\bibnamefont {Zoller}}, \bibinfo {author} {\bibfnamefont
  {A.~A.}\ \bibnamefont {Clerk}},\ and\ \bibinfo {author} {\bibfnamefont
  {L.}~\bibnamefont {Jiang}},\ }\href@noop {} {\bibinfo {title} {Autonomous
  quantum error correction and fault-tolerant quantum computation with squeezed
  cat qubits}} (\bibinfo {year} {2022}{\natexlab{a}}),\ \Eprint
  {https://arxiv.org/abs/arXiv:2210.13406} {arXiv:2210.13406} \BibitemShut
  {NoStop}%
\bibitem [{\citenamefont {Michael}\ \emph {et~al.}(2016)\citenamefont
  {Michael}, \citenamefont {Silveri}, \citenamefont {Brierley}, \citenamefont
  {Albert}, \citenamefont {Salmilehto}, \citenamefont {Jiang},\ and\
  \citenamefont {Girvin}}]{Michael2016}%
  \BibitemOpen
  \bibfield  {author} {\bibinfo {author} {\bibfnamefont {M.~H.}\ \bibnamefont
  {Michael}}, \bibinfo {author} {\bibfnamefont {M.}~\bibnamefont {Silveri}},
  \bibinfo {author} {\bibfnamefont {R.~T.}\ \bibnamefont {Brierley}}, \bibinfo
  {author} {\bibfnamefont {V.~V.}\ \bibnamefont {Albert}}, \bibinfo {author}
  {\bibfnamefont {J.}~\bibnamefont {Salmilehto}}, \bibinfo {author}
  {\bibfnamefont {L.}~\bibnamefont {Jiang}},\ and\ \bibinfo {author}
  {\bibfnamefont {S.~M.}\ \bibnamefont {Girvin}},\ }\bibfield  {title}
  {\bibinfo {title} {New class of quantum error-correcting codes for a bosonic
  mode},\ }\href {https://doi.org/10.1103/PhysRevX.6.031006} {\bibfield
  {journal} {\bibinfo  {journal} {Phys. Rev. X}\ }\textbf {\bibinfo {volume}
  {6}},\ \bibinfo {pages} {031006} (\bibinfo {year} {2016})}\BibitemShut
  {NoStop}%
\bibitem [{Ft1()}]{Ft1}%
  \BibitemOpen
  \href@noop {} {}\bibinfo {note} {We modified the GKP codes
  \cite{Gottesman2001} to be rotationally invariant.}\BibitemShut {Stop}%
\bibitem [{Ft2()}]{Ft2}%
  \BibitemOpen
  \href@noop {} {}\bibinfo {note} {In the present work, we focus on exploiting
  the rotation-invariance of bosonic codes which is compatible with the
  cavity-QED-based approach of \cite{Li2023}. Additional translation-invariance
  is useful to enhance the loss robustness of the codes
  \cite{Gottesman2001,Schlegel2022,Xu2022}, but will not be considered here. We
  also do not consider concatenations of the rotation-invariant bosonic codes
  into higher-level codes with one logical qubit composed of many qubits, using
  here only single-mode bosonic codes and thus keeping the strongest notion of
  hardware efficiency.}\BibitemShut {Stop}%
\bibitem [{\citenamefont {Mirrahimi}\ \emph {et~al.}(2014)\citenamefont
  {Mirrahimi}, \citenamefont {Leghtas}, \citenamefont {Albert}, \citenamefont
  {Touzard}, \citenamefont {Schoelkopf}, \citenamefont {Jiang},\ and\
  \citenamefont {Devoret}}]{Mirrahimi2014}%
  \BibitemOpen
  \bibfield  {author} {\bibinfo {author} {\bibfnamefont {M.}~\bibnamefont
  {Mirrahimi}}, \bibinfo {author} {\bibfnamefont {Z.}~\bibnamefont {Leghtas}},
  \bibinfo {author} {\bibfnamefont {V.~V.}\ \bibnamefont {Albert}}, \bibinfo
  {author} {\bibfnamefont {S.}~\bibnamefont {Touzard}}, \bibinfo {author}
  {\bibfnamefont {R.~J.}\ \bibnamefont {Schoelkopf}}, \bibinfo {author}
  {\bibfnamefont {L.}~\bibnamefont {Jiang}},\ and\ \bibinfo {author}
  {\bibfnamefont {M.~H.}\ \bibnamefont {Devoret}},\ }\bibfield  {title}
  {\bibinfo {title} {Dynamically protected cat-qubits: a new paradigm for
  universal quantum computation},\ }\href
  {https://doi.org/10.1088/1367-2630/16/4/045014} {\bibfield  {journal}
  {\bibinfo  {journal} {New Journal of Physics}\ }\textbf {\bibinfo {volume}
  {16}},\ \bibinfo {pages} {045014} (\bibinfo {year} {2014})}\BibitemShut
  {NoStop}%
\bibitem [{\citenamefont {Bergmann}\ and\ \citenamefont {van
  Loock}(2016)}]{Bergmann2016}%
  \BibitemOpen
  \bibfield  {author} {\bibinfo {author} {\bibfnamefont {M.}~\bibnamefont
  {Bergmann}}\ and\ \bibinfo {author} {\bibfnamefont {P.}~\bibnamefont {van
  Loock}},\ }\bibfield  {title} {\bibinfo {title} {Quantum error correction
  against photon loss using multicomponent cat states},\ }\href
  {https://doi.org/10.1103/PhysRevA.94.042332} {\bibfield  {journal} {\bibinfo
  {journal} {Phys. Rev. A}\ }\textbf {\bibinfo {volume} {94}},\ \bibinfo
  {pages} {042332} (\bibinfo {year} {2016})}\BibitemShut {NoStop}%
\bibitem [{\citenamefont {Leghtas}\ \emph {et~al.}(2013)\citenamefont
  {Leghtas}, \citenamefont {Kirchmair}, \citenamefont {Vlastakis},
  \citenamefont {Schoelkopf}, \citenamefont {Devoret},\ and\ \citenamefont
  {Mirrahimi}}]{Leghtas2013}%
  \BibitemOpen
  \bibfield  {author} {\bibinfo {author} {\bibfnamefont {Z.}~\bibnamefont
  {Leghtas}}, \bibinfo {author} {\bibfnamefont {G.}~\bibnamefont {Kirchmair}},
  \bibinfo {author} {\bibfnamefont {B.}~\bibnamefont {Vlastakis}}, \bibinfo
  {author} {\bibfnamefont {R.~J.}\ \bibnamefont {Schoelkopf}}, \bibinfo
  {author} {\bibfnamefont {M.~H.}\ \bibnamefont {Devoret}},\ and\ \bibinfo
  {author} {\bibfnamefont {M.}~\bibnamefont {Mirrahimi}},\ }\bibfield  {title}
  {\bibinfo {title} {Hardware-efficient autonomous quantum memory protection},\
  }\href {https://doi.org/10.1103/PhysRevLett.111.120501} {\bibfield  {journal}
  {\bibinfo  {journal} {Phys. Rev. Lett.}\ }\textbf {\bibinfo {volume} {111}},\
  \bibinfo {pages} {120501} (\bibinfo {year} {2013})}\BibitemShut {NoStop}%
\bibitem [{\citenamefont {Stoler}(1970)}]{Stoler1970}%
  \BibitemOpen
  \bibfield  {author} {\bibinfo {author} {\bibfnamefont {D.}~\bibnamefont
  {Stoler}},\ }\bibfield  {title} {\bibinfo {title} {Equivalence classes of
  minimum uncertainty packets},\ }\href
  {https://doi.org/10.1103/PhysRevD.1.3217} {\bibfield  {journal} {\bibinfo
  {journal} {Phys. Rev. D}\ }\textbf {\bibinfo {volume} {1}},\ \bibinfo {pages}
  {3217} (\bibinfo {year} {1970})}\BibitemShut {NoStop}%
\bibitem [{\citenamefont {Lu}(1972)}]{Lu1972}%
  \BibitemOpen
  \bibfield  {author} {\bibinfo {author} {\bibfnamefont {E.~Y.~C.}\
  \bibnamefont {Lu}},\ }\bibfield  {title} {\bibinfo {title} {Quantum
  correlations in two-photon amplification},\ }\href
  {https://doi.org/10.1007/BF02762058} {\bibfield  {journal} {\bibinfo
  {journal} {Lettere al Nuovo Cimento (1971-1985)}\ }\textbf {\bibinfo {volume}
  {3}},\ \bibinfo {pages} {585} (\bibinfo {year} {1972})}\BibitemShut {NoStop}%
\bibitem [{\citenamefont {Hollenhorst}(1979)}]{Hollenhorst1979}%
  \BibitemOpen
  \bibfield  {author} {\bibinfo {author} {\bibfnamefont {J.~N.}\ \bibnamefont
  {Hollenhorst}},\ }\bibfield  {title} {\bibinfo {title} {Quantum limits on
  resonant-mass gravitational-radiation detectors},\ }\href
  {https://doi.org/10.1103/PhysRevD.19.1669} {\bibfield  {journal} {\bibinfo
  {journal} {Phys. Rev. D}\ }\textbf {\bibinfo {volume} {19}},\ \bibinfo
  {pages} {1669} (\bibinfo {year} {1979})}\BibitemShut {NoStop}%
\bibitem [{Ft3()}]{Ft3}%
  \BibitemOpen
  \href@noop {} {}\bibinfo {note} {Our definition of the codewords of squeezed
  cat codes allows to improve the success probability of discriminating between
  the codewords because the overlap of these codewords is minimized along the
  direction orthogonal to the polar direction. This is distinct from the
  definitions in \cite{Schlegel2022} and \cite{xu2022b}, where the components
  of the squeezed cat states are squeezed along the polar
  direction.}\BibitemShut {Stop}%
\bibitem [{\citenamefont {Gottesman}\ \emph {et~al.}(2001)\citenamefont
  {Gottesman}, \citenamefont {Kitaev},\ and\ \citenamefont
  {Preskill}}]{Gottesman2001}%
  \BibitemOpen
  \bibfield  {author} {\bibinfo {author} {\bibfnamefont {D.}~\bibnamefont
  {Gottesman}}, \bibinfo {author} {\bibfnamefont {A.}~\bibnamefont {Kitaev}},\
  and\ \bibinfo {author} {\bibfnamefont {J.}~\bibnamefont {Preskill}},\
  }\bibfield  {title} {\bibinfo {title} {Encoding a qubit in an oscillator},\
  }\href {https://doi.org/10.1103/PhysRevA.64.012310} {\bibfield  {journal}
  {\bibinfo  {journal} {Phys. Rev. A}\ }\textbf {\bibinfo {volume} {64}},\
  \bibinfo {pages} {012310} (\bibinfo {year} {2001})}\BibitemShut {NoStop}%
\bibitem [{\citenamefont {Albert}\ \emph {et~al.}(2018)\citenamefont {Albert},
  \citenamefont {Noh}, \citenamefont {Duivenvoorden}, \citenamefont {Young},
  \citenamefont {Brierley}, \citenamefont {Reinhold}, \citenamefont {Vuillot},
  \citenamefont {Li}, \citenamefont {Shen}, \citenamefont {Girvin},
  \citenamefont {Terhal},\ and\ \citenamefont {Jiang}}]{Albert2018}%
  \BibitemOpen
  \bibfield  {author} {\bibinfo {author} {\bibfnamefont {V.~V.}\ \bibnamefont
  {Albert}}, \bibinfo {author} {\bibfnamefont {K.}~\bibnamefont {Noh}},
  \bibinfo {author} {\bibfnamefont {K.}~\bibnamefont {Duivenvoorden}}, \bibinfo
  {author} {\bibfnamefont {D.~J.}\ \bibnamefont {Young}}, \bibinfo {author}
  {\bibfnamefont {R.~T.}\ \bibnamefont {Brierley}}, \bibinfo {author}
  {\bibfnamefont {P.}~\bibnamefont {Reinhold}}, \bibinfo {author}
  {\bibfnamefont {C.}~\bibnamefont {Vuillot}}, \bibinfo {author} {\bibfnamefont
  {L.}~\bibnamefont {Li}}, \bibinfo {author} {\bibfnamefont {C.}~\bibnamefont
  {Shen}}, \bibinfo {author} {\bibfnamefont {S.~M.}\ \bibnamefont {Girvin}},
  \bibinfo {author} {\bibfnamefont {B.~M.}\ \bibnamefont {Terhal}},\ and\
  \bibinfo {author} {\bibfnamefont {L.}~\bibnamefont {Jiang}},\ }\bibfield
  {title} {\bibinfo {title} {Performance and structure of single-mode bosonic
  codes},\ }\href {https://doi.org/10.1103/PhysRevA.97.032346} {\bibfield
  {journal} {\bibinfo  {journal} {Phys. Rev. A}\ }\textbf {\bibinfo {volume}
  {97}},\ \bibinfo {pages} {032346} (\bibinfo {year} {2018})}\BibitemShut
  {NoStop}%
\bibitem [{\citenamefont {Bennett}\ \emph {et~al.}(1992)\citenamefont
  {Bennett}, \citenamefont {Brassard},\ and\ \citenamefont
  {Mermin}}]{Bennett1992}%
  \BibitemOpen
  \bibfield  {author} {\bibinfo {author} {\bibfnamefont {C.~H.}\ \bibnamefont
  {Bennett}}, \bibinfo {author} {\bibfnamefont {G.}~\bibnamefont {Brassard}},\
  and\ \bibinfo {author} {\bibfnamefont {N.~D.}\ \bibnamefont {Mermin}},\
  }\bibfield  {title} {\bibinfo {title} {Quantum cryptography without bell's
  theorem},\ }\href {https://doi.org/10.1103/PhysRevLett.68.557} {\bibfield
  {journal} {\bibinfo  {journal} {Phys. Rev. Lett.}\ }\textbf {\bibinfo
  {volume} {68}},\ \bibinfo {pages} {557} (\bibinfo {year} {1992})}\BibitemShut
  {NoStop}%
\bibitem [{\citenamefont {Scarani}\ \emph {et~al.}(2009)\citenamefont
  {Scarani}, \citenamefont {Bechmann-Pasquinucci}, \citenamefont {Cerf},
  \citenamefont {Du\ifmmode~\check{s}\else \v{s}\fi{}ek}, \citenamefont
  {L\"utkenhaus},\ and\ \citenamefont {Peev}}]{Scarani2009}%
  \BibitemOpen
  \bibfield  {author} {\bibinfo {author} {\bibfnamefont {V.}~\bibnamefont
  {Scarani}}, \bibinfo {author} {\bibfnamefont {H.}~\bibnamefont
  {Bechmann-Pasquinucci}}, \bibinfo {author} {\bibfnamefont {N.~J.}\
  \bibnamefont {Cerf}}, \bibinfo {author} {\bibfnamefont {M.}~\bibnamefont
  {Du\ifmmode~\check{s}\else \v{s}\fi{}ek}}, \bibinfo {author} {\bibfnamefont
  {N.}~\bibnamefont {L\"utkenhaus}},\ and\ \bibinfo {author} {\bibfnamefont
  {M.}~\bibnamefont {Peev}},\ }\bibfield  {title} {\bibinfo {title} {The
  security of practical quantum key distribution},\ }\href
  {https://doi.org/10.1103/RevModPhys.81.1301} {\bibfield  {journal} {\bibinfo
  {journal} {Rev. Mod. Phys.}\ }\textbf {\bibinfo {volume} {81}},\ \bibinfo
  {pages} {1301} (\bibinfo {year} {2009})}\BibitemShut {NoStop}%
\bibitem [{\citenamefont {Abruzzo}\ \emph {et~al.}(2013)\citenamefont
  {Abruzzo}, \citenamefont {Bratzik}, \citenamefont {Bernardes}, \citenamefont
  {Kampermann}, \citenamefont {van Loock},\ and\ \citenamefont
  {Bru\ss{}}}]{Abruzzo2013}%
  \BibitemOpen
  \bibfield  {author} {\bibinfo {author} {\bibfnamefont {S.}~\bibnamefont
  {Abruzzo}}, \bibinfo {author} {\bibfnamefont {S.}~\bibnamefont {Bratzik}},
  \bibinfo {author} {\bibfnamefont {N.~K.}\ \bibnamefont {Bernardes}}, \bibinfo
  {author} {\bibfnamefont {H.}~\bibnamefont {Kampermann}}, \bibinfo {author}
  {\bibfnamefont {P.}~\bibnamefont {van Loock}},\ and\ \bibinfo {author}
  {\bibfnamefont {D.}~\bibnamefont {Bru\ss{}}},\ }\bibfield  {title} {\bibinfo
  {title} {Quantum repeaters and quantum key distribution: Analysis of
  secret-key rates},\ }\href {https://doi.org/10.1103/PhysRevA.87.052315}
  {\bibfield  {journal} {\bibinfo  {journal} {Phys. Rev. A}\ }\textbf {\bibinfo
  {volume} {87}},\ \bibinfo {pages} {052315} (\bibinfo {year}
  {2013})}\BibitemShut {NoStop}%
\bibitem [{\citenamefont {Kirby}\ \emph {et~al.}(2016)\citenamefont {Kirby},
  \citenamefont {Santra}, \citenamefont {Malinovsky},\ and\ \citenamefont
  {Brodsky}}]{Kirby2016}%
  \BibitemOpen
  \bibfield  {author} {\bibinfo {author} {\bibfnamefont {B.~T.}\ \bibnamefont
  {Kirby}}, \bibinfo {author} {\bibfnamefont {S.}~\bibnamefont {Santra}},
  \bibinfo {author} {\bibfnamefont {V.~S.}\ \bibnamefont {Malinovsky}},\ and\
  \bibinfo {author} {\bibfnamefont {M.}~\bibnamefont {Brodsky}},\ }\bibfield
  {title} {\bibinfo {title} {Entanglement swapping of two arbitrarily degraded
  entangled states},\ }\href {https://doi.org/10.1103/PhysRevA.94.012336}
  {\bibfield  {journal} {\bibinfo  {journal} {Phys. Rev. A}\ }\textbf {\bibinfo
  {volume} {94}},\ \bibinfo {pages} {012336} (\bibinfo {year}
  {2016})}\BibitemShut {NoStop}%
\bibitem [{\citenamefont {Wu}\ \emph {et~al.}(1986)\citenamefont {Wu},
  \citenamefont {Kimble}, \citenamefont {Hall},\ and\ \citenamefont
  {Wu}}]{Wu1986}%
  \BibitemOpen
  \bibfield  {author} {\bibinfo {author} {\bibfnamefont {L.-A.}\ \bibnamefont
  {Wu}}, \bibinfo {author} {\bibfnamefont {H.~J.}\ \bibnamefont {Kimble}},
  \bibinfo {author} {\bibfnamefont {J.~L.}\ \bibnamefont {Hall}},\ and\
  \bibinfo {author} {\bibfnamefont {H.}~\bibnamefont {Wu}},\ }\bibfield
  {title} {\bibinfo {title} {Generation of squeezed states by parametric down
  conversion},\ }\href {https://doi.org/10.1103/PhysRevLett.57.2520} {\bibfield
   {journal} {\bibinfo  {journal} {Phys. Rev. Lett.}\ }\textbf {\bibinfo
  {volume} {57}},\ \bibinfo {pages} {2520} (\bibinfo {year}
  {1986})}\BibitemShut {NoStop}%
\bibitem [{\citenamefont {{Lo Franco}}\ \emph {et~al.}(2010)\citenamefont {{Lo
  Franco}}, \citenamefont {Compagno}, \citenamefont {Messina},\ and\
  \citenamefont {Napoli}}]{LoFranco2010}%
  \BibitemOpen
  \bibfield  {author} {\bibinfo {author} {\bibfnamefont {R.}~\bibnamefont {{Lo
  Franco}}}, \bibinfo {author} {\bibfnamefont {G.}~\bibnamefont {Compagno}},
  \bibinfo {author} {\bibfnamefont {A.}~\bibnamefont {Messina}},\ and\ \bibinfo
  {author} {\bibfnamefont {A.}~\bibnamefont {Napoli}},\ }\bibfield  {title}
  {\bibinfo {title} {Efficient generation of n-photon binomial states and their
  use in quantum gates in cavity qed},\ }\href
  {https://doi.org/https://doi.org/10.1016/j.physleta.2010.03.036} {\bibfield
  {journal} {\bibinfo  {journal} {Physics Letters A}\ }\textbf {\bibinfo
  {volume} {374}},\ \bibinfo {pages} {2235} (\bibinfo {year}
  {2010})}\BibitemShut {NoStop}%
\bibitem [{\citenamefont {Miry}\ \emph {et~al.}(2014)\citenamefont {Miry},
  \citenamefont {Tavassoly},\ and\ \citenamefont {Roknizadeh}}]{Miry2014}%
  \BibitemOpen
  \bibfield  {author} {\bibinfo {author} {\bibfnamefont {S.~R.}\ \bibnamefont
  {Miry}}, \bibinfo {author} {\bibfnamefont {M.~K.}\ \bibnamefont
  {Tavassoly}},\ and\ \bibinfo {author} {\bibfnamefont {R.}~\bibnamefont
  {Roknizadeh}},\ }\bibfield  {title} {\bibinfo {title} {On the generation of
  number states, their single- and two-mode superpositions, and two-mode
  binomial state in a cavity},\ }\href
  {https://doi.org/10.1364/JOSAB.31.000270} {\bibfield  {journal} {\bibinfo
  {journal} {J. Opt. Soc. Am. B}\ }\textbf {\bibinfo {volume} {31}},\ \bibinfo
  {pages} {270} (\bibinfo {year} {2014})}\BibitemShut {NoStop}%
\bibitem [{\citenamefont {Chuang}\ \emph {et~al.}(1997)\citenamefont {Chuang},
  \citenamefont {Leung},\ and\ \citenamefont {Yamamoto}}]{Chuang1997}%
  \BibitemOpen
  \bibfield  {author} {\bibinfo {author} {\bibfnamefont {I.~L.}\ \bibnamefont
  {Chuang}}, \bibinfo {author} {\bibfnamefont {D.~W.}\ \bibnamefont {Leung}},\
  and\ \bibinfo {author} {\bibfnamefont {Y.}~\bibnamefont {Yamamoto}},\
  }\bibfield  {title} {\bibinfo {title} {Bosonic quantum codes for amplitude
  damping},\ }\href {https://doi.org/10.1103/PhysRevA.56.1114} {\bibfield
  {journal} {\bibinfo  {journal} {Phys. Rev. A}\ }\textbf {\bibinfo {volume}
  {56}},\ \bibinfo {pages} {1114} (\bibinfo {year} {1997})}\BibitemShut
  {NoStop}%
\bibitem [{\citenamefont {Xu}\ \emph {et~al.}(2022{\natexlab{b}})\citenamefont
  {Xu}, \citenamefont {Zheng}, \citenamefont {Wang}, \citenamefont {Zoller},
  \citenamefont {Clerk},\ and\ \citenamefont {Jiang}}]{Xu2022}%
  \BibitemOpen
  \bibfield  {author} {\bibinfo {author} {\bibfnamefont {Q.}~\bibnamefont
  {Xu}}, \bibinfo {author} {\bibfnamefont {G.}~\bibnamefont {Zheng}}, \bibinfo
  {author} {\bibfnamefont {Y.-X.}\ \bibnamefont {Wang}}, \bibinfo {author}
  {\bibfnamefont {P.}~\bibnamefont {Zoller}}, \bibinfo {author} {\bibfnamefont
  {A.~A.}\ \bibnamefont {Clerk}},\ and\ \bibinfo {author} {\bibfnamefont
  {L.}~\bibnamefont {Jiang}},\ }\href
  {https://doi.org/10.48550/ARXIV.2210.13406} {\bibinfo {title} {Autonomous
  quantum error correction and fault-tolerant quantum computation with squeezed
  cat qubits}} (\bibinfo {year} {2022}{\natexlab{b}})\BibitemShut {NoStop}%
\end{thebibliography}%
\end{document}